\documentclass[twocolumn,showpacs,preprintnumbers,amsmath,amssymb]{revtex4}

\usepackage{graphicx}
\usepackage{dcolumn}
\usepackage{bm}
\newcommand{\bi}[1]{\mbox{\boldmath$#1$}}
\newcommand{\av}[1]{\langle{#1}\rangle}
\renewcommand{\theequation}{\arabic{section}.\arabic{equation}}
\def\be{\begin{equation}}
\def\en{\end{equation}} 
\def\p{\partial }  
\def\ve{\varepsilon}
\def\gs{\gtrsim}
\def\ls{\lesssim}
\def\bphi{\bar{\phi}}

\def\bea{\begin{eqnarray}}
\def\ena{\end{eqnarray}}

\begin{document}
\preprint{APS}
\title{Precipitation  in 
aqueous mixtures  with addition of strongly 
hydrophilic or hydrophobic solute }

\author{Ryuichi Okamoto}
\author{Akira Onuki}
\affiliation{Department of Physics, Kyoto University, Kyoto 606-8502, Japan}

\date{\today}

\begin{abstract}
We examine phase separation in aqueous   mixtures 
due to  preferential solvation 
with   a low-density solute 
(hydrophilic  ions   or hydrophobic particles). 
For hydrophilic ions, 
preferential solvation 
can stabilize water  domains enriched with ions.  
 This precipitation 
 occurs   in  wide ranges of the temperature  
and the average composition above a critical 
solute  density $n_{\rm p}$, where  
the  mixture solvent would be  in a one-phase 
state without solute.  The volume fraction of 
precipitated   domains  tends   to zero 
as the average solute  density $\bar n$ 
is decreased  to   $n_{\rm p}$ or 
as the interaction parameter 
$\chi$ is decreased to a  critical value $\chi_{\rm p}$. 
If we start with  one-phase states with 
 ${\bar n}>n_{\rm p}$ or $\chi>\chi_{\rm p}$, 
 precipitation proceeds  via homogeneous nucleation 
or via   heterogeneous nucleation, for example,  around 
 suspended colloids. In the latter case, colliod particles are 
 wrapped  by thick wetting layers. 
We also predict a first-order prewetting transition  
 for $\bar n$ or $\chi$ 
 slightly below $n_{\rm p}$ or $\chi_{\rm p}$.  
\end{abstract}


\maketitle

%
%
 
\section{Introduction}

In  fluid mixtures  composed of  water and 
a  less polar organic liquid, 
phase separation behavior 
can be drastically changed by  a small amount of  
salt \cite{polar1,polar3,Misawa,Kumar,cluster,Taka,Anisimov,po}.  
With addition of a 
10$^{-3}$ mole fraction of hydrophilic salt 
like NaCl,  the coexistence curve  is  typically shifted 
 by a few Kelvins  
 upward in  the USCT case or downward in  the LSCT case.  
In some aqueous  mixtures, 
even  if they are  miscible at all $T$ 
at atmosphere pressure without salt, addition of  
a  small amount of  salt 
gives  rise to reentrant phase separation behavior 
 \cite{Kumar,cluster,Anisimov,Misawa}.

With ions added,  moreover, many  
groups have also found   long-lived 
heterogeneities (sometimes extending  over 
a few micrometers)  in one-phase states 
\cite{So} and a  third phase 
visible  as a thin solid-like 
plate  at  a liquid-liquid interface 
in two-phase states \cite{third}. A representative system is a  
 mixture of H$_2$O (or D$_2$O),  tri-methylpyridine (3MP), 
and  NaBr.  The heterogeneities  have been detected  
by dynamic light scattering \cite{SCAT}, indicating  
 spontaneous formation of 
   aggregates or domains with small diffusion constants. 
Their size and the volume fraction are 
$50 {\rm \AA}$  and $10^{-6}$ as a typical  example. 
 The third phase have been observed also 
in isobutylic acid (IBA)-water 
mixtures without added salt, where 
the isobutyric acid partly 
dissociates into Butyrate$^-$ and H$^+$  ions. 
In addition, a drastic decrease in 
the ion mobility was observed in  IBA-water  
for high contents  of IBA \cite{Bonn}.  
Recently, mesophases have been observed  
when    sodium tetraphenylborate 
NaBPh$_{4}$ was added to    D$_2$O-3MP 
at about $ 0.1$ M \cite{Sadakane}.  
This salt dissociates into  
 hydrophilic  Na$^+$  and hydrophobic  BPh$_{4}^-$.  
The latter ion  consists  of  four phenyl rings bonded 
to an ionized boron.

Dramatic  ion effects  are ubiquitous 
in various soft materials such as polymers, gels, colloids, 
and mixtures containing  ionic surfactants. 
For example,  complex effects are  known 
to be induced in polyelectrolytes  
when a second fluid component (cosolvent) 
is added to   water. 
In particular, precipitation of DNA has been widely observed 
with addition of alcohol such as ethanol to water 
\cite{Bloomfield}.  
Here  the alcohol added  is 
excluded from condensed DNA, which suggests 
solvation-induced wetting of 
DNA by water.   We also mention observations 
of  crystal formation of 
micron-sized water-rich 
droplets in a less polar oil-rich phase  
 with addition of HBr \cite{Bla}. The mechanism 
 of this ordering was 
 ascribed to the Coulomb interaction 
 due to asymmetric partitioning 
 of cations and anions in  water and oil \cite{Roij}.

In these phenomena, 
the  solvation (or hydration) 
 interaction   among 
ions and polar  molecules should
play a major role. Small hydrophilic ions  
are  solvated by several water molecules  \cite{Is,Osakai,Wakisaka} and 
the resultant solvation chemical potential per ion  
much exceeds the thermal energy $k_BT$ and strongly 
depends on the ambient composition of the solvent. 
On the contrary, some 
hydrophobic  particles 
dislike to be in contact 
with water molecules, which  tend to form 
aggregates in water and are 
more soluble in oil than in water.  
  In  aqueous  mixtures, 
a hydrophobic ion should be  surrounded by 
oil molecules, which 
explains the  mesophase   formation 
observed by  Sadakane {\it et al} \cite{Sadakane}.
Hydrophilic and hydrophobic particles 
 strongly affect the surrounding hydrogen bonding network  
 in different manners \cite{Is} and  the 
solvation is  highly cooperative.  
Furthermore,   when a small amount of 
water was added to 
methanol-cyclohexane,  the coexistence 
curve  was largely shifted \cite{Jacobs}  
and a water-induced, methanol-rich  
wetting layer appeared (which was nonexistent without water)  
\cite{Be}. 
These findings  indicate strongly  selective 
 molecular  interactions   between  water and 
 the two components.

Obviously,   the preferential solvation 
should strongly affect the phase transition behavior  in 
aqueous mixtures with a small amount of 
 hydrophilic or hydrophobic 
solute, though   
this aspect has not been well studied.  
Recently,  some  efforts have been made 
to elucidate the solvation effects 
in phase separation  in 
 mixture solvents   in electrolytes 
\cite{Onuki-Kitamura,OnukiPRE,Tsori,Roij1,Andelman1,Araki,Daan}, 
polyelectrolytes\cite{Okamoto}, 
and ionic surfactants \cite{OnukiEPL}. 
A review on the static properties 
was presented \cite{OnukiReview}. In the dynamics, a 
number of  problems still remain unexplored   
 \cite{Araki,Daan}.

In this paper, we present 
a theory of solvation-induced phase separation 
with  hydrophilic or hydrophobic solute. 
Thus we aim to  
explain the observed heterogeneities in aqueous 
mixtures \cite{So}. 
We may  suppose   hydrophilic 
monovalent ion pairs such as Na$^+$ and Cl$^-$ 
in  a binary mixture  of water and a less polar 
component (called oil). As is well-known, 
hydrophilic ions    induce 
clustering of  water molecules around them 
on microscopic scales, forming 
a solvation shell  \cite{Is,Osakai,Wakisaka}.  
We shall see  that 
strongly hydrophilic ions  can  moreover induce  formation 
 of water-rich domains on macroscopic scales 
for sufficiently  strong preference  of water over oil. 
This can occur even when the mixture is 
outside the coexistence curve without ions.

The organization of this paper is as follows. 
In Sec.II,   
two-phase coexistence induced by  
the preferential solvation will be studied 
numerically and theoretically, where  
the electrostatic interaction does not appear explicitly.   
In Sec.III, inhomogeneous profiles such as interfaces 
 will be calculated. We will show that a precipitated droplet 
 can be stable only above a minimum radius 
 due to the surface tension effect 
 and that a solvation-induced prewetting 
 transition occurs far from the coexistence curve without 
 solute. In Sec.IV, two-phase coexistence with 
hydrophilic ions will be examined in the presence of 
the charge effect. 
In Sec.V, we will investigate 
the precipitation  from  one-phase states 
taking place as homogeneous nucleation. 
In Sec.IV, we will investigate adsorption and 
precipitation on  colloid surfaces.

\section{Phase separation with  strongly selective solute}
\setcounter{equation}{0}

Neglecting electrostatic interaction but accounting for the 
solvation interaction, we first 
consider  a  binary mixture composed 
of water and a less polar component 
in a cell with a  fixed volume $V$. 
The second component will be simply called oil 
hereafter. Ions will be treated as 
a dilute third component (solute) with density $n({\bi r})$. 
We may also suppose strongly hydrophilic or hydrophobic 
neutral particles (possibly with a complex structure) 
 added in an aqueous binary mixture. 
Such  particles can be  solvated  by 
a certain number of water or oil molecules  depending 
on whether they are  hydrophilic or hydrophobic.

In the following, we will suppose hydrophilic ions 
or particles. However, our results can be equally applicable to 
 hydrophobic ions or  particles  
if  water and oil are 
exchanged (or  $\phi$ is replaced by $1-\phi$). 
The  Boltzmann constant $k_B$ 
will be  set equal to unity.

\subsection{Conditions of two phase coexistence with solute }

The volume fractions 
of water,  oil, and solute    are  written as 
$\phi({\bi r})$ and  $\phi'({\bi r})$, and 
$v_I n({\bi r})$, respectively, where $v_I$ is the solute   
 volume. If the two solvent species  have the same 
molecular volume $v_0$, their densities are 
$\phi/v_0$ and  $\phi'/v_0$.  The solvent diameter 
is of order $a=v_0^{1/3}$.  The 
 space-filling  condition is written as  
\be   
\phi+\phi'+ v_I n=1.  
\en 
The solute volume fraction 
is assumed to be  small or 
$v_I n\ll 1$. This is more    easily satisfied  
if the solute size is smaller 
than that of solvent or  $v_I < v_0$. 
In  this paper, 
to simplify   the calculations, we thus set   
\be 
\phi'=1-\phi. 
\en

The  free energy density $f_{\rm tot}(\phi,n)$ 
for the composition $\phi$ and the solute density $n$ 
consists of three parts as   
\be
{f}_{\rm tot} (\phi,n)  =  f(\phi) +T{n}\ln (n\lambda^3) 
 - Tg n\phi.    
\en 
The first term  is assumed to be 
of the Bragg-Williams  form for a fluid mixture 
\cite{Onukibook}, 
\be 
{f}(\phi)  =  \frac{T}{v_0}[\phi \ln\phi + (1-\phi)\ln (1-\phi) 
+ \chi \phi (1-\phi)],  
\en
where  $\chi$ is the interaction parameter 
 dependent  on  $T$. The length $a=v_0^{1/3}$ 
 represents the molecular diameter ($\sim 3 {\rm \AA}$ for water). 
In Eq.(2.3),  the second term arises 
from  the solute translational entropy.  
The $\lambda$ is the thermal de Broglie length 
(but the terms linear in $n$ are irrelevant 
and $\lambda^3$ may be replaced by $v_0$ 
 in the following).  The third term arises from 
the solute  preference  of  water over oil. 
The parameter      
$g$  is assumed to much exceed unity ($\gg 1$)\cite{OnukiPRE}.   
We  fix  the average densities  
of the constituent components, which are expressed as  
\be 
 \bphi = \int d{\bi r} \phi/V, \quad 
{\bar n}= \int d{\bi r} n/V. 
\en 
In our theory, $\chi$, $\bar\phi$, and $\bar n$ 
are relevant control parameters. 
Changing $\chi$ through the critical value 
$\chi_c=2$ is equivalent to changing $T$ through 
the solvent critical temperature $T_c$.

In   two phase coexistence in equilibrium, let 
the composition and the solute density be 
$(\phi_\alpha,n_\alpha)$ and  $(\phi_\beta,n_\beta)$ 
in  phases $\alpha$ and $\beta$ with $\phi_\beta<{\bar \phi}
<\phi_\alpha$ and $n_\beta<{\bar n}
<n_\alpha$.  Here we 
 give simple thermodynamic arguments. 
We define the  chemical potentials 
as $h= {\p f_{\rm tot}}/{\p \phi}$ and 
 $\mu= {\p f_{\rm tot}}/{\p n}$. From Eq.(2.3) we obtain 
\bea 
h&=&  
f'(\phi) -Tgn, \\
\mu
&=& T[\ln (n\lambda^3)+1- g\phi],
\ena 
where $f'=\p f/\phi$. 
First, the homogeneity 
relation $\mu(\phi_\alpha,n_\alpha)=\mu(\phi_\beta,n_\beta)$  
for solute gives rise to  
\be 
n_\alpha= A_0 e^{g\phi_\alpha},  
\quad n_\beta= A_0  e^{g\phi_\beta}. 
\en 
The  coefficient $A_0=\lambda^{-3}\exp(\mu/T-1)$ is 
determined from the conservation of the solute 
number in Eq.(2.5)  as 
\be 
A_0= {\bar n} /\av{e^{g\phi}}, 
\en 
where $\av{\cdots}$ denotes   the space average in the cell. 
Let   $\gamma_\alpha$ be   
the volume fraction of 
 phase $\alpha$. Neglecting the volume of the 
 interface region,  we obtain 
\be 
\av{e^{g\phi}}= \gamma_\alpha e^{g\phi_\alpha} 
 +(1-\gamma_\alpha) e^{g\phi_\beta}
\en 
From Eq.(2.5) $\gamma_\alpha$  is expressed in terms of 
$\bar\phi$ and $\bar n$  as 
\bea 
\gamma_\alpha &=& ({{\bar \phi}-\phi_\beta})/{\Delta \phi}
\\
&=&  ({{\bar n}-n_\beta})/{\Delta n}.  
\ena  
The differences between the two phases are written as  
\be 
\Delta \phi=\phi_\alpha-\phi_\beta>0, \quad 
 \Delta n=n_\alpha-n_\beta>0.
\en   
With the aid of Eq.(2.8), the ratio 
 $n_\alpha/{\bar n}$ is written in terms 
 of the compositions   as  
\be 
n_\alpha/{\bar n}= \Delta\phi/[
({\bar\phi}-\phi_\beta)+
(\phi_\alpha-{\bar\phi})e^{-g\Delta\phi}].
\en  
For not small $\Delta \phi$, we find 
 $n_\alpha/n_\beta = e^{g\Delta\phi}\gg 1$ and $\Delta n
 \cong n_\alpha$.  Namely,     
the solute density is much higher in 
 phase $\alpha$ than in phase $\beta$.  
The solvation part $-Tgn_\alpha\phi$  
of  the  free energy density  
is  significant in phase 
$\alpha$ even for small $\bar n$.

Next, the homogeneity relation $h(\phi_\alpha,n_\alpha)=
h(\phi_\beta,n_\beta)$ for the solvent composition  is written as 
\bea 
 h&=&f'(\phi_\alpha) -Tgn_\alpha\nonumber\\
 &=& f'(\phi_\beta) -Tgn_\beta.    
\ena   
In equilibrium, we also  require minimization 
of the grand potential density $\omega$ 
defined by    
\bea 
\omega &=& f_{\rm tot}- h\phi-\mu n \nonumber\\
&=&  f- h\phi -T n ,
\ena 
where the second line 
follows from Eq.(2.7). In the two phases,  $\omega$ 
assumes the same value so that 
\be 
{f(\phi_\alpha)-f(\phi_\beta)}- T\Delta n=h{\Delta\phi}. 
\en

We may derive Eqs.(2.15) and (2.17) 
from minimization of 
the total free energy $F$ under Eq.(2.5). 
Using  $\gamma_\alpha$ we  express $F$  as 
\be 
{F}/{V}= \gamma_\alpha f_{\rm tot}(\phi_\alpha,n_\alpha) 
+ (1-\gamma_\alpha) f_{\rm tot}(\phi_\beta,n_\beta) ,
\en  
where the surface free energy is neglected. 
If use is made of the relation 
$n(\ln (n \lambda^3)-g\phi)= n \ln A_0$ in the two phases, 
the above expression is rewritten as 
\bea 
{F}/{V}
&=& \gamma_\alpha f(\phi_\alpha) 
+ (1-\gamma_\alpha) f(\phi_\beta)+ T{\bar n}
 \ln ({\bar n}\lambda^3)   \nonumber \\
&-& T{\bar n} \ln [ 
 \gamma_\alpha e^{g\phi_\alpha} 
 +(1-\gamma_\alpha) e^{g\phi_\beta}] .
\ena 
The third term in the first line is a constant 
at constant $\bar{n}= \av{n}$  
and is irrelevant, but 
the fourth term in the second line is a singular 
solvation contribution   
at fixed $\bar{n}= \av{n}$ 
relevant for large $g$ (even for small $\bar n$). 
Since $\bar{\phi}= \av{\phi}$ is also fixed,  
 we should minimize 
\be 
{\tilde\omega}=  F/V-h[\gamma_\alpha \phi_\alpha  
 +(1-\gamma_\alpha) \phi_\beta-{\bar\phi}]
\en 
with respect to $\phi_\alpha$, 
$\phi_\beta$, and $\gamma_\alpha$.  
Here  $h$ appears as the Lagrange multiplier. 
 With the aid of  the expression (2.19), 
 the minimum conditions,   $\p {\tilde\omega}/\p \phi_\alpha= 
\p {\tilde\omega}/\p \phi_\beta= \p {\tilde\omega}/\p \gamma_\alpha=0$, 
readily lead  to    Eqs.(2.15) and  (2.17).

Without solute ${\bar n}=0$, 
two-phase coexistence is possible only for $\chi>2$. 
Thus, in the appendix A, we  will perform  
the Taylor expansions  
of $\phi_\alpha$ and $\phi_\beta$ with respect to $\bar n$ 
for $\chi>2$.  We shall see 
that  the water-rich coexistence branch 
is much deformed even by a very small amount 
of highly preferential solute  
(as can be seen in Fig.1 below).

\subsection{Numerical results of 
precipitation}

\begin{figure}[t]
\begin{center}
\includegraphics[scale=0.49]{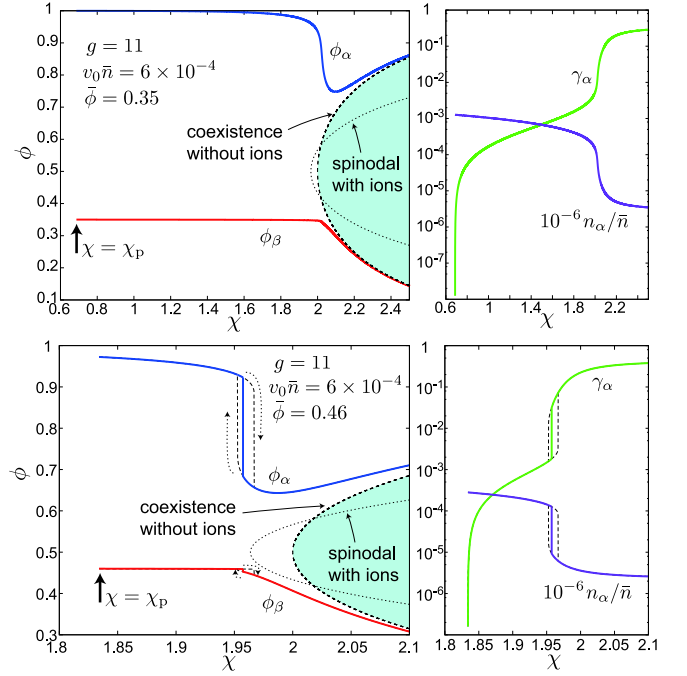}
\caption{(Color online) 
Left:Compositions $\phi_\alpha$ and $\phi_\beta$ vs $\chi$. 
Right:Semi-logarithmic plots of 
volume fraction  $\gamma_\alpha$  
and normalized ion density 
$n_\alpha/{\bar n}$   of the water-rich phase $\alpha$ vs $\chi$.   
Here  ${\bar n}= 6\times 10^{-4}v_0^{-1}$  and $g=11$. 
Precipitation occurs for $ \chi_{\rm p}<\chi\ls 2$.  
For $\bar{\phi}= 0.35$ (top), $\phi_\alpha$ changes continuously. 
For $\bar{\phi}= 0.46$ (bottom), 
$\phi_\alpha$ jumps at $\chi \cong 2$.  Plotted 
in the left  also are the coexisting  region without solute (in blue) 
and the spinodal curve with solute (see Eqs.(3.7) and 
(5.2))  \cite{OnukiPRE}.}
\end{center}
\end{figure}

\begin{figure}[th]
\begin{center}
\includegraphics[scale=0.5]{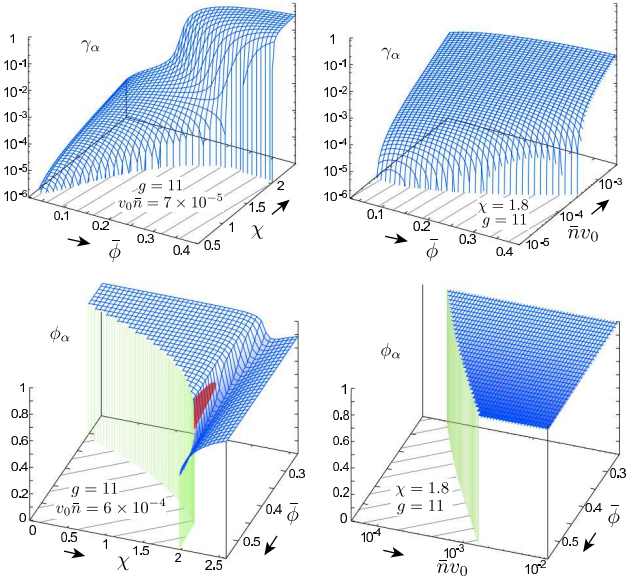}
\caption{(Color online) Volume fraction  
 $\gamma_\alpha$ (top) 
and composition $\phi_\alpha$ (bottom) 
of the water-rich phase $\alpha$ 
in the $\chi$-$\bar \phi$ plane at fixed $\bar n$ (left) 
and in the $\bar n$-$\bar \phi$ plane at fixed $\chi$ (right), 
where $g=11$.  Precipitation occurs 
for $\chi>\chi_{\rm p}({\bar \phi},{\bar n})$ (left) and  
${\bar n}> n_{\rm p}({\bar \phi},{\chi})$ (right) outside 
 the slashed regions. 
In the left bottom,    
a first-order transition 
occurs for  $\chi\cong 1.95$ and ${\bar \phi}>0.395$ 
 (in red). See  the bottom plates in Fig.1. }
\end{center}
\end{figure}

\begin{figure}[t]
\begin{center}
\includegraphics[scale=0.45]{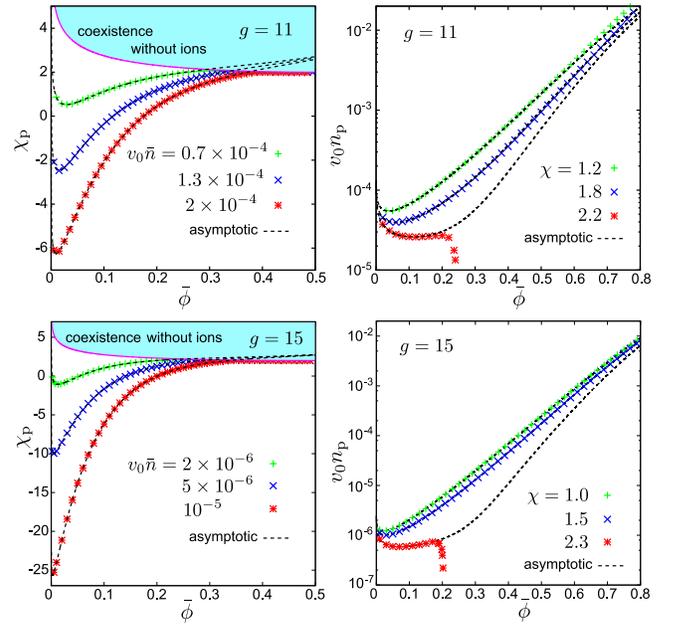}
\caption{(Color online) Left:$\chi_{\rm p}(\bphi,{\bar n})$ 
vs $\bar \phi$ for three values of $\bar n$ 
at $g=11$ (top) and 15 (bottom), 
which nearly coincide with the asymptotic formula (2.33) 
(dotted line) for $\bar{\phi} <0.35$ and 
converge to the coexistence curve for larger $\bar\phi$. 
Coexistence region  without  ions is also shown (in blue).
Right: $v_0 n_{\rm p}(\bphi,\chi)$ 
vs $\bar \phi$  on  a semi-logarithmic scale 
for three values of $\chi$ at $g=11$ (top) and 15 (bottom), 
It coincides with the asymptotic formula (2.34) 
(dotted line) for $\chi<2$ and  tends  to zero at  
the coexistence composition,  0.249   (top)  
or  0.204  (bottom). 
}
\end{center}
\end{figure}

Next we give  numerical results 
 on  the phase behavior of our system. 
 We will set $g=11$  mostly,  
 but will also  present 
 additional  results for $g=15$ in Fig.3 below.
The solute density $n$ will be measured in units of $v_0^{-1}$.

In Fig.1, we display   $\phi_\alpha$ and  
 $\phi_\beta$ in the left panels  
 and  $n_\alpha$ and 
  $\gamma_\alpha$ in the right panels  
  as functions of $\chi$ (or $T$ experimentally) 
  at  ${\bar n}= 6\times 10^{-4}v_0^{-1}$. 
Due to  the nonlinear solute effect, 
  a precipitation branch appears in the range, 
\be  
\chi_{\rm p} ({\bar \phi},{\bar n}) < \chi< 2,
\en  
where the solvent would be in one-phase states without solute. 
Here $\gamma_\alpha$ 
tends to zero as $\chi \to 
\chi_{\rm p}$, where   the lower bound  
$\chi_{\rm p}$ depends  on $\bar\phi$ and $\bar n$. 
This branch appears  under the condition 
 $e^{g(1-\bar{\phi})}\gg 1$ (see the next 
subsection).      For  large $g(=11$ here), 
the  precipitated domains become 
solute-rich (salted water for hydrophilic ions) with 
\be 
\phi_\alpha \cong 1, \quad  
 n_\alpha\gg  n_\beta=e^{-g\Delta\phi}n_\alpha.
\en   
The right panels are on  semi-logarithmic scales.  
With decreasing $\chi$, we can see 
an increase in the solute density 
$n_\alpha$ and an decrease in the  
volume fraction $\gamma_\alpha$. 
Details of the figure are as follows. 
(i) For    $\bphi=0.35$ (top), 
  $\phi_\alpha$ changes rather 
  abruptly around $\chi \sim 2$ 
but continuously as a function of $\chi$, being 
 minimum at $\chi=2.05$. The 
  $n_\alpha$ increases up to $
0.381v_0^{-1}$ as $\chi\to \chi_{\rm p}=0.687$. 
(ii) For   $\bphi=0.46$ (bottom), 
the precipitation branch much shrinks  
with $\chi_{\rm p}=1.834$, while    
$n_\alpha= 0.0842v_0^{-1}$ at 
$\chi=\chi_{\rm p}$. For this larger $\bar\phi$, 
 $\phi_\alpha$ changes discontinuously on  a hysteresis 
loop in the range $1.953<\chi<1.967$. 
If  $F$ in Eq.(2.18) is  minimized, 
a  first-order   transition  is found to occur     at 
$\chi=1.957$.  Even for $\chi>2$,  
the two-phase coexistence is much 
deformed by the solute.

In Fig.2, we show  $\gamma_\alpha$  
and $\phi_\alpha$ in the $\chi$-$\bar \phi$ 
plane at  ${\bar n}= 6\times 10^{-4}v_0^{-1}$  (left) 
and in the $\bar n$-$\bar \phi$ plane at  $\chi=1.8$ (right). 
For fixed $\chi$ in the $\bar n$-$\bar \phi$ plane, 
precipitation occurs for 
\be  
 {\bar n}> n_{\rm p}({\bar \phi},{\chi}),
\en   
where  $\gamma_\alpha$ tends to zero as ${\bar n} \to 
n_{\rm p}$. This  minimum solute density 
$n_{\rm p}$ depends on $\bar\phi$ and $\chi$.   
As the crossover to the asymptotic behavior 
 in Eq.(2.22), 
$\phi_\alpha$ becomes  appreciably smaller 
than $1$ for   $\chi \sim 2$ or for ${\bar \phi} \sim 0.5$. 
For large $\bar\phi\gs 0.5$,  the  precipitation 
branch shrinks to vanish for small solute densities, where   
water molecules are 
already abundant around solute 
particles in one-phase states.

In Fig.3,      at $g=11$ and 15, 
we show $\chi_{\rm p}(\bphi, {\bar n})$ 
vs $\bar \phi$   
for three values of $\bar n$  (left) 
and $n_{\rm p}(\bphi,\chi)$ vs  $\bar \phi$ 
for three values of  $\chi$ (right). 
These quantities decrease 
dramatically at small $\bar\phi$ and 
their magnitudes strongly depend on  $g$. 
Even for $\chi>2$,  $n_{\rm p}$ can be calculated 
outside the water-rich branch of the 
coexistence curve and it tends to zero as 
$\bar\phi$ approaches the coexistence 
composition without solute. 
Theoretical explanations of 
their behavior will be given in the following 
subsections.

\subsection{Theory of asymptotic behavior}

We present  a theory on  the asymptotic behavior of  
the precipitation branch 
for $g\gg 1$ in the region $\chi<2$. 
At its starting point, we assume 
 the branch satisfying Eq.(2.22)  
and confirm its existence 
 self-consistently. Since $\phi_\alpha\cong 1$, 
 the logarithmic term $(\propto (1-\phi)\ln(1-\phi)$) 
 in the free energy density (2.4)  is crucial 
 in phase $\alpha$.

We first  neglect the term   $-Tgn_\beta$ in Eq.(2.15) 
from $gv_0 n_\beta \ll 1$ and the term 
$f(\phi_\alpha)$ 
in Eq.(2.17) from 
$f(\phi_\alpha) 
\sim T(1-\phi_\alpha) \ln (1-\phi_\alpha)$. Note that 
$gv_0 n_\beta\sim 10^{-2}$ in Fig.1. 
It follows a simplified 
equation,   
\be 
h \cong   
f'(\phi_\beta)\cong 
- [{f(\phi_\beta)+Tn_\alpha}]/({1-\phi_\beta}) . 
\en 
This determines the  solute density $n_\alpha$ 
in  phase $\alpha$ 
as a function of $\phi_\beta$ in the form,  
\be 
n_\alpha\cong G(\phi_\beta)/T,
\en 
where $G(\phi)$ is  defined by  
\bea 
G(\phi) &=& -f(\phi) - (1-\phi)f'(\phi)\nonumber\\
&=& - (T/v_0) [\ln \phi+ \chi(1-\phi)^2].
\ena 
Using the free energy 
(2.4), we obtain the second line 
and plot it in the $\phi$-$\chi$ plane in Fig.4. 
It  is positive outside the 
coexistence curve, ensuring $n_\alpha>0$ in Eq.(2.25).

\begin{figure}[t]
\begin{center}
\includegraphics[scale=0.6]{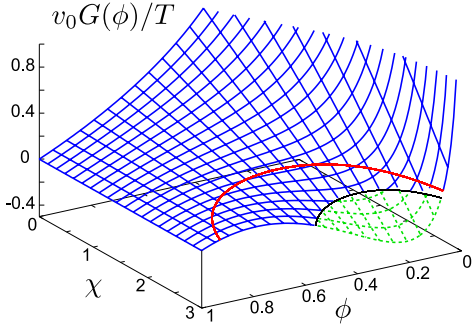}
\caption{(Color online) $v_0 G(\phi)/T$ defined in Eq.(2.26) 
in the  $\phi$-$\chi$ plane. 
It is negative only in the inner green region, 
where the minimum of $\chi$ is 
2.455 at $\phi=0.285$. 
It is positive outside it in the blue region.
The coexistence curve without solute is written 
in red, on which the minimum of $\chi$ is 2 at $\phi=0.5$. 
}
\end{center}
\end{figure}

In Eq.(2.15),   we  next 
use Eq.(2.4) for $f'(\phi_\alpha)$ to obtain 
\be 
h  \cong v_0^{-1}
 T [-\ln (1-\phi_\alpha)-\chi]-Tg n_\alpha\cong f'(\phi_\beta),
\en 
  where  the logarithmic term ($\propto \ln(1-\phi)$) 
balances with the solvation term   ($\propto gn_\alpha$).  
Use of  Eq.(2.26) gives    
\be 
 {1-\phi_\alpha}\cong A_\beta \exp[-g v_0G(\phi_\beta)/T],
\en 
where the coefficient $A_\beta$  is given by 
\be 
A_\beta=\exp[-\chi - v_0 f'(\phi_\beta)/T],
\en 
so $A_\beta$ is of order unity.  
The factor $\exp[-g v_0 G(\phi_\beta)/T]$  in Eq.(2.28) 
is very small for $g\gg 1$, leading to $\phi_\alpha \cong 1$ 
 as in Figs.1 and 2.

Furthermore, from Eqs. (2.12) and (2.25), 
the volume fraction $\gamma_\alpha$ 
of phase $\alpha$ is approximated as  
\be 
\gamma_\alpha 
\cong 
{\bar n}/n_\alpha -e^{-g\Delta \phi}\cong 
T{\bar n}/G(\phi_\beta) -e^{-g\Delta \phi}.
\en   
The above relation is rewritten as 
\bea  
&&\hspace{-1.2cm}G(\phi_\beta) \cong  {T{\bar n}}/{
(\gamma_\alpha+  e^{-g\Delta\phi})}\nonumber\\
&&\hspace{-2mm}\cong\frac{T{\bar n}(1-\phi_\beta)}{
\bar{\phi}-\phi_\beta+  (1-\phi_\beta)
\exp[{-g(1-\phi_\beta)]}}. 
\ena 
From the first to  second line,  we have used   Eq.(2.11)  and 
 replaced  $\Delta\phi$ by $ 1-\phi_\beta$. This 
 equation also  follows from Eqs.(2.14) and (2.25). 
It  determine $\phi_\beta$ and   $\gamma_\alpha
\cong ({\bar \phi}-\phi_\beta)/(1-\phi_\beta)$.  
 We  recognize that  
 $G(\phi_\beta)$ increases  up to $T{\bar n}e^{g\Delta\phi}
\cong  T{\bar n}e^{g(1-{\bar\phi})}$  as  
$\gamma_\alpha \to 0$ or as $\phi_\beta \to \bar\phi$. 
In this  limit  it follows   the marginal relation,
\be 
G({\bar \phi})\cong  T{\bar n}
e^{g(1-{\bar\phi})} \quad (\gamma_\alpha\to 0).
\en 
If ${\bar n}$ is fixed above $n_{\rm p}$, 
Eq.(2.32)  holds 
at ${\chi}=\chi_{\rm p}$ so that 
\be 
\chi_{\rm p} \cong 
 [-\ln \bphi- v_0{\bar n}e^{g(1-\bphi)}]/(1-\bphi)^2,  
\en 
where we use the second line of  Eq.(2.26). 
Notice that the solute density 
$\bar n$ appears in the combination ${\bar n}e^{g(1-\bphi)}$ 
and its effect of lowering  $\chi_{\rm p}$ 
is much amplified for  $g(1-\phi)\gg 1$ 
even for very small $\bar n$. 
(For ${\bar n}=0$ the right hand side of Eq.(2.33) 
is above the coexistence curve.) 
On the other hand, if $\chi$ is fixed above $\chi_{\rm p}$, Eq.(2.32)  holds 
at ${\bar n}=n_{\rm p}$. Thus the minimum solute density 
$n_{\rm p}$ is estimated as 
\be 
n_{\rm p} \cong  e^{-g(1-{\bar \phi})}G({\bar \phi})/T, 
\en
which  is much decreased 
by the factor $e^{-g(1-{\bar \phi})}$.

In Fig.3, the  curves of $\chi_{\rm p}$ and $n_{\rm p}$ 
 nearly coincide with  the 
 asymptotic formulas (2.33) and (2.34) in the range 
 ${\bar\phi}<0.35$ 
for $\chi_{\rm p}$  and  in the range 
 $\chi<2$ for ${n}_{\rm p}$. They  exhibit a minimum 
 at $\bphi \sim 
e^{-g}/v_0{\bar n}g$ for $\chi_{\rm p}$ 
and  at $\bphi \sim 
g^{-1}$ for ${n}_{\rm p}$. For larger $\bar{\phi}>0.35$,  
$\chi_{\rm p}$ nearly  coincide with the coexistence curve, 
indicating  disappearance of the precipitation branch. 
 Notice that   ${n}_{\rm p}$ 
decreases to   zero as $\bar\phi$ approaches 
the coexistence composition  $\phi_{\rm cx}
= 0.249$   at  $\chi=2.2$ (top)  
and 0.204  at  $\chi=2.3$ (bottom), 
where phase separation occurs without solute.

\subsection{Theory for small volume fraction of precipitates}

We next construct an  analytic theory 
of  precipitation  for small volume 
fraction  $\gamma_\alpha \ll 1$. 
We will calculate $F$ in Eq.(2.19) 
treating  $\gamma_\alpha$ as an order parameter 
of the phase transition. Though this theory 
 is applicable only for  $\gamma_\alpha \ll 1$, 
it can yield  Eq.(2.21) in the asymptotic 
limit. 

In Eq.(2.19) we first expand 
$\phi_\beta$ and $f(\phi_\beta)$ 
 with respect to $\gamma_\alpha$ as 
\bea 
&&\hspace{-2mm} \phi_\beta =({\bar \phi}-\phi_\alpha)/(1-\gamma_\alpha)
 \cong   {\bar \phi}+
 ({\bar\phi}-\phi_\alpha)\gamma_\alpha, \nonumber\\
&&\hspace{-4mm}f(\phi_\beta) \cong  
f(\bphi) + f'({\bar\phi}) 
(\bphi- \phi_\alpha) \gamma_\alpha .
\ena   
Let $F_0= Vf_{\rm tot}({\bar \phi},{\bar n})$ be 
the one-phase value of $F$ (at $\gamma_\alpha=0$). 
The increment $\Delta F=F-F_0$ is 
due to precipitate formation and is of the form,  
\be 
{\Delta F}/{V}=   \Omega(\phi_\alpha) 
 \gamma_\alpha - T {\bar n} \ln 
 [1+ \Psi(\phi_\alpha)  \gamma_\alpha ] ,
\en   
where $ \omega_b(\phi)$ and $\Psi(\phi)$ are defined by 
\bea 
&&
\Omega(\phi) =f(\phi)- f({\bar\phi}) 
- f' ({\bar \phi})(\phi-\bar{\phi}),\\ 
&& 
\Psi (\phi) = \exp[{g(\phi-{\bar\phi})}]-1-
 g(\phi-{\bar\phi}), 
\ena  
where $\bar\phi$ is treated as a constant and its 
dependence of $\Omega$ and $\Phi$ are suppressed. 
In Eq.(2.36) we  have not expanded the logarithmic term 
with respect to $ \gamma_\alpha$,  because 
 the coefficient  
$\Psi(\phi_\alpha)$ grows strongly 
as $e^{g(\phi_\alpha-\bar{\phi})}$  
for $g(\phi_\alpha-\bar{\phi})\gg 1$. 
For $\chi<2$,  the positivity   $\Omega(\phi_\alpha)>0$  
follows in the region ${\bar\phi}<\phi<\phi_\alpha$  
since Eq.(2.37) gives 
\be 
\Omega(\phi_\alpha)= \int_{\bar \phi}^{\phi_\alpha}
d\phi  (\phi_\alpha- \phi) f''(\phi) ,  
\en 
where  $f''(\phi) = \p^2 f/\p^2\phi>0$. 
Note that if $\phi$ and $\bar \phi$ in 
$\Omega(\phi)$ are replaced by $1$ and $\phi_\beta$, 
respectively, $\Omega(\phi)$ becomes identical to  
$G(\phi_\beta)$ in Eq.(2.26). 
If use is made of Eq.(2.4), 
 $\Omega(\phi)$ is explicitly written as  
\be 
\frac{v_0}{T}\Omega(\phi) = \phi \ln (\frac{\phi}{{\bar\phi}}) +
(1-\phi ) \ln (\frac{1-\phi}{1-{\bar\phi}})
 -\chi (\phi-{\bar\phi})^2. 
\en 

\begin{figure}[t]
\begin{center}
\includegraphics[scale=0.5]{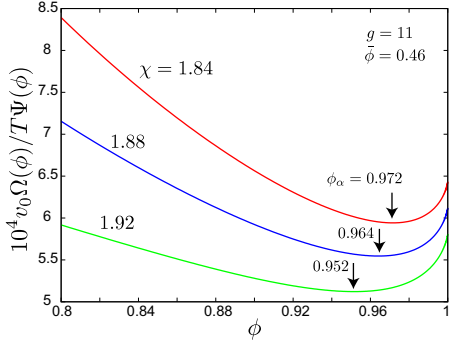}
\caption{(Color online) 
$\Omega(\phi)/\Psi(\phi)$ multiplied by 
$10^4v_0/T$  for three values of $\chi$ at 
 $g=11$ and ${\bar\phi}=0.46$. 
Each curve exhibits a minimum at $\phi=\phi_\alpha$, 
where $\phi_\alpha$ is estimated as in  Eq.(2.51). 
Each minimum value is equal to $10^4v_0n_{\rm p}$ from Eq.(2.46).}
\end{center}
\end{figure}

\begin{figure}[t]
\begin{center}
\includegraphics[scale=0.45]{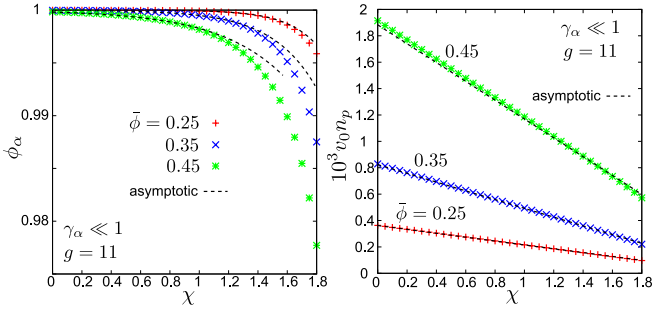}
\caption{(Color online) 
 Theoretical $\phi_\alpha$ in the range $[0.975,1]$ (left) 
 and  $n_{\rm p}$ multiplied by $10^3v_0$ (right) 
for  ${\bar \phi}=0.25,0.35$, and 0.45 
 at $g=11$.  
These curves are calculated for  $\gamma_\alpha \ll 1$ 
from Eqs.(2.45) and (2.46). For these parameters, 
they only slightly 
deviate from the asymptotic formulas 
 in Eqs.(2.49) and  (2.52). 
   The curves of 
$n_{\rm p}$  are nearly linear in $\chi$ 
from the  linear dependence of  
$G({\bar\phi})$ on $\chi$.}
\end{center}
\end{figure}
 
For each given $\bar \phi$ and $\bar n$, 
$\Delta F$ should be minimized 
 with respect to $\gamma_\alpha$ and 
$\phi_\alpha$. Here $\phi_\beta$ is  given by Eq.(2.35). 
From Eq.(2.36) its derivative with respect to $\gamma_\alpha$ 
is  
\be 
\frac{1}{V} \frac{\p{\Delta F} }{\p\gamma_\alpha}=   \Omega(\phi_\alpha) 
 - \frac{T {\bar n} \Psi(\phi_\alpha)}{
1+ \Psi(\phi_\alpha)  \gamma_\alpha } .
\en 
The right hand side can vanish only when 
\be 
Z(\phi_\alpha) \equiv \Omega(\phi_\alpha)/T{\bar n}\Psi(\phi_\alpha)<1.
\en 
If  $Z(\phi_\alpha)<1$, 
$\gamma_\alpha$ is nonvanishing and is written as  
\be 
\gamma_\alpha = T{\bar n}/\Omega(\phi_\alpha) 
- 1/\Psi(\phi_\alpha).
\en 
For  this   $\gamma_\alpha$,  $\Delta F$ 
in Eq.(2.36) becomes negative as   
\be 
{\Delta F}/{V}=  T{\bar n}[1-Z(\phi_\alpha) +\ln Z(\phi_\alpha)] .
\en   
In the range $0<Z<1$, 
the function $1-Z+\ln Z$ is negative and 
decreases with decreasing $Z$. 
In order to  minimize $\Delta F$,  we should thus minimize  
 $Z(\phi)$  or  the function 
$\Omega(\phi)/\Psi(\phi)$ at $\phi=\phi_\alpha$.  
The equation to determine 
$\phi_\alpha=\phi_\alpha({\bar \phi},\chi)$ is hence  given by 
\be 
\frac{\p}{\p \phi_\alpha} \frac{\Omega(\phi_\alpha)}{\Psi(\phi_\alpha)}=
 \frac{\Omega'(\phi_\alpha)}{\Psi(\phi_\alpha)}-
\frac{\Omega(\phi_\alpha)}{\Psi(\phi_\alpha)^2}\Psi'(\phi_\alpha)
=0,
\en 
where $\Omega'(\phi)=\p \Omega(\phi)/\p \phi$ and 
 $\Psi'(\phi)=\p \Psi(\phi)/\p \phi$. 
Note that $\phi_\alpha$ is independent of $\bar n$ for $\gamma_\alpha \ll 1$. 
In Fig.5, we plot  $\Omega(\phi)/\Psi(\phi)$ 
for three values of $\chi$ at ${\bar\phi}= 0.46$
 to demonstrate the existence of 
 its  minimum at $\phi=\phi_\alpha$. 
In the left panel of Fig.6, we show 
$\phi_\alpha$  calculated from the above equation.

In terms of  the above $\phi_\alpha$ 
(dependent on $\bar \phi$ and $\chi$),  
the  minimum solute density $n_{\rm p}$  is 
obtained  from the condition $Z(\phi_\alpha)=1$ as 
\be 
n_{\rm p}= \Omega(\phi_\alpha)/T\Psi(\phi_\alpha),
\en 
which is a function of $\bar \phi$ and $\chi$. 
Then $Z(\phi_\alpha)=n_{\rm p}/{\bar n}$. 
The condition  $Z(\phi_\alpha)<1$ 
yields ${\bar n}>n_{\rm p}$, as ought to be the case, 
and  $\gamma_\alpha$  is also expressed as      
\be 
\gamma_\alpha  =  ({\bar n}/n_{\rm p}-1)/\Psi(\phi_\alpha).
\en 
In the right panel of Fig.6, we show 
 $n_{\rm p}$ calculated from Eq.(2.46) using the data of $\phi_\alpha$ 
in the left panel.

After  $\phi_\alpha$ and $n_{\rm p}$ have been  determined, 
Eq.(2.47) indicates that 
an arbitrary value can be assigned to 
  ${\bar n}$  in the range 
$0<\bar{n}/n_{\rm p}-1 \ll \Psi(\phi_\alpha)$.
Furthermore, by  decreasing   $\chi$ at fixed  $\bar n$ and 
 $\bar \phi$, we can make 
$Z(\phi_\alpha)$ approach unity, where 
$\chi \to \chi_{\rm p}=\chi_{\rm p}({\bar \phi},\bar{n})$ in Eq.(2.21). 
Using Eq.(2.40) we may express $\chi_{\rm p}$ as    
\bea 
\chi_{\rm p}&=& \bigg[
\phi_\alpha \ln \frac{\phi_\alpha}{{\bar\phi}}+
(1-\phi_\alpha ) \ln \frac{1-\phi_\alpha}{1-{\bar\phi}}
\bigg]/(\phi_\alpha-{\bar\phi})^2\nonumber\\
&&
 -v_0{\bar n}\Psi(\phi_\alpha) /(\phi_\alpha-{\bar\phi})^2. 
\ena 
The relations (2.46) and (2.48) for the lower bounds are  exact  
in our model and are  consistent with the   asymptotic 
ones  (2.33) and (2.34) 
in the limit  of $\phi_\alpha \cong 1$ and $\Psi (\phi_\alpha)
\cong e^{g(1-\phi)}\gg 1$.

 We now give analysis of the function 
 $\Omega(\phi)/\Psi(\phi)$ in the range $\phi<{\bar \phi}$.  
  If $\phi$   is  not very close to unity, 
$\Omega(\phi)/\Psi(\phi)$   decreases rapidly with increasing $\phi$,  
because $1/\Psi(\phi) \propto e^{-g\phi}$ decreases rapidly. 
However, as $\phi \to 1$, the logarithmic term 
in $f(\phi)$ eventually comes into play. 
To show this, we  approximate  $\Omega(\phi)$ 
for   $\phi \cong  1$  as 
\be 
{\Omega(\phi)}\cong G({\bar\phi})+ \frac{T}{v_0}(1-\phi) 
 \bigg[ \ln\frac{1-\phi}{A_0}-1\bigg],
\en 
where $G(\phi)$ is defined in Eq.(2.26) 
and  $\ln(A_0)=-\chi- v_0f'({\bar \phi})/T$. 
The coefficient  $A_\beta$ in Eq.(2.29) 
tends to $A_0$ as $\phi_\beta \to {\bar \phi}$.  The derivative of 
 $\Omega(\phi)/\Psi(\phi)$ with respect to $\phi$ 
 is thus estimated as 
\be 
 \frac{d}{d\phi} \frac{\Omega(\phi)}{\Psi(\phi)} 
 \cong \bigg[ -g G({\bar\phi})- \frac{T}{v_0}
 \ln\frac{1-\phi}{A_0} 
\bigg] e^{-g (\phi-{\bar \phi})}.
\en 
In this derivative, 
 the logarithmic term grows weakly  but can  
 balance the solvation term ($\propto g$) as $\phi\to 1$, 
 so that 
\be 
1-\phi_\alpha\cong A_0 \exp(-gv_0G({\bar\phi})/T)
\en 
in accord with Eq.(2.28). 
The asymptotic form of $n_{\rm p}$ is 
written as in Eq.(2.34) if we 
set $\Omega(\phi_\alpha) 
\cong G({\bar\phi})$ and  $\Psi(\phi) \cong  e^{g\Delta\phi}$ 
in  Eq.(2.46).

\section{Inhomogeneous composition profiles}

Here, still  neglecting 
 the charge effect,  
we examine the surface tension, 
 the stability of  a spherical droplet,   and 
 the  surface adsorption  near a boundary wall.    
These problems stem from the 
 solvation-induced phase separation.

\subsection{Gradient free energy and surface tension}
\setcounter{equation}{0}

 Including  the gradient free energy 
 we  assume the following simple 
 form for the total free  energy,
\be 
F= \int d{\bi r}[{ f}_{\rm tot} (\phi,n)  
+ \frac{C}{2}|\nabla \phi|^2],
\en  
where  $ f_{\rm tot}$ is given by Eq.(2.3) 
and $C$ is assumed to be a positive constant. 
We suppose  a planar interface at $z=0$ 
varying along the $z$ axis 
and separating two phases $\alpha$ and $\beta$. Then 
$\phi=\phi(z)$ and $n=n(z)$  
are functions of  $z$ 
 with $\phi(-\infty)=\phi_\alpha$ 
and $\phi(\infty)=\phi_\beta$. In equilibrium, 
the functional derivatives 
of $F$ with respect to $\phi$ and $n$ are 
constant in space. Therefore, 
\be
h= \delta F/\delta\phi=   
f'(\phi)- Tg n -C \phi''={\rm constant},
\en
where $\phi''= d^2\phi/dz^2$. Since 
$\mu= \delta F/\delta n= \p f_{\rm tota}/\p n$ is given by Eq.(2.7), 
the solute density depends on $\phi$ as  
\be 
n(z) =A_0\exp[{g\phi(z)}],
\en 
where the coefficient $A_0$ is determined by Eq.(2.9) at 
fixed ${\bar n}= \av{n}$.   

We multiply  Eq.(3.2)   by $\phi'= d\phi/dz$ 
and integrate it  over $z$ to obtain 
\be 
\Delta \omega(\phi)= 
\omega(\phi)-\omega(\phi_\beta)= \frac{1}{2}{C}({\phi}')^2,
\en  
where $\omega(\phi) =f -T n-h \phi$ 
is the grand potential density in  Eq.(2.16) 
satisfying $\omega(\phi_\alpha)= \omega(\phi_\beta)$ as 
in  Eq.(2.17). From Eq.(2.15) and the second line of 
Eq.(2.16), the  difference  
 $\Delta \omega(\phi)$ is  written as  
\bea 
&&\hspace{-1cm}\Delta \omega(\phi)
= f(\phi)- f(\phi_\beta)-f'(\phi_\beta)(\phi-\phi_\beta)
\nonumber\\   
&&- Tn_\beta[ e^{g(\phi-\phi_\beta) }- g(\phi-\phi_\beta)-1]. 
\ena 
It is worth noting that 
the limit  $\gamma_\alpha \to 0$ 
can be attained at  ${\bar n}=n_{\rm p}$.  
From Eq.(3.5) we find 
\be 
\lim_{\gamma_\alpha \to 0} \Delta\omega(\phi) 
=  \Omega (\phi)-
[{\Omega (\phi_\alpha)}/{\Psi (\phi_\alpha)}]\Psi(\phi), 
\en  
where $\Omega(\phi)$ and   $\Psi(\phi)$ are defined in Eqs.(2.37) 
and (2.38).  
From Eq.(2.45) the derivative of the left hand side 
of Eq.(3.6) with respect to $\phi$ 
vanishes at $\phi= \phi_\alpha$. 
There is no singular behavior in the interface profile 
even in the limit  $\gamma_\alpha \to 0$. 

The  relation $\Delta\omega(\phi)\cong 
 \omega''(\phi_\beta) (\phi-\phi_\beta)^2/2$  
for  $\phi \cong  \phi_\beta$ follows directly from Eq.(3.5). 
The $\Delta\omega(\phi)$ also behaves as  
$ \omega''(\phi_\alpha)  (\phi- \phi_\alpha)^2/2$  
for  $\phi \cong  \phi_\alpha$  
from Eqs.(2.15) and (2.17). 
Here  $ \omega''(\phi)
= f''(\phi) -Tg^2 n(\phi)$ is the second derivative 
of $\Delta\omega(\phi)$ 
with respect to $\phi$ and the correlation lengths  
$\xi_\alpha$ and $\xi_\beta$ 
in the bulk two phases are  given by 
\bea 
C\xi_\alpha^{-2} &=& f''(\phi_\alpha) -Tg^2 n_\alpha,
\nonumber\\ 
C\xi_\beta^{-2} &=& f''(\phi_\beta) -Tg^2 n_\beta, 
\ena 
in terms of which  
$\phi(z)-\phi_\alpha \propto e^{-|z|/\xi_\alpha}$ for 
$z < -\xi_\alpha$ and 
$\phi(z)-\phi_\beta \propto e^{-z/\xi_\beta}$ for 
$z>  \xi_\beta$. 
The right hand sides of Eq.(3.7) should be  positive 
for the stability of the bulk two phases. 
However,  since $f''(\phi_\alpha)\sim  T/v_0(1-\phi_\alpha)$, 
$\xi_\alpha$ becomes even shorter than $a=v_0^{1/3}$ 
for $\phi_\alpha\cong 1$ and  $C\sim T/a$. 
Thus  the composition 
 varies too steeply  in phase $\alpha$  in the 
present gradient theory. However, 
we shall see  (around Eq.(3.11) below)  
that  the surface tension is 
mainly determined by the composition variation  in 
the phase $\beta$ side 
of an  interface.

In $\Delta\omega(\phi)$ in Eq.(3.5), 
 the  last term in the second line 
grows as $\exp[g(\phi-\phi_\beta)]$ 
for $\phi -\phi_\beta\gg 1/g$ 
and can  balance the other  terms 
with increasing $\phi$  even for very small $n_\beta$. 
 In Fig.7, we  illustrates  this behavior by 
plotting  the sum of the  first three terms 
and the minus of the last term 
in the right hand side of 
Eq.(3.5) separately. That is, we  write 
 $\Delta\omega(\phi) = 
\Delta\omega_0(\phi) -\omega_{\rm sol}(\phi)$ with   
\bea
&&\hspace{-6mm}\Delta\omega_0(\phi) = 
f(\phi)- f(\phi_\beta)-f'(\phi_\beta)(\phi-\phi_\beta), 
\nonumber\\
&&\hspace{-6mm}\omega_{\rm sol}(\phi) =   
Tn_\beta [ e^{g(\phi-\phi_\beta) }- g(\phi-\phi_\beta)-1] . 
\ena   
The  $\Delta\omega_0(\phi)$ 
is the grand potential difference 
without solute. As a function of $\phi$, it increases   
 monotonically for  $\chi<2$   
and exhibits     positive extrema  
for $\chi>2$ (with $\phi_\beta$ 
being outside the coexistence curve).  

\begin{figure}[t]
\begin{center}
\includegraphics[scale=0.45]{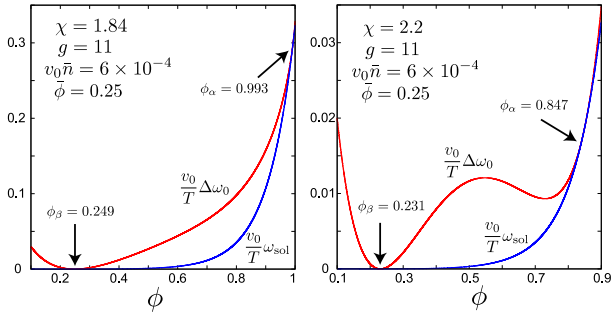}
\caption{
$\Delta\omega_0(\phi)$ 
and  $\omega_{\rm sol}(\phi)$ in eq.(2.59) 
in units of $T/v_0$ for $\chi=1.84$ (left) 
and $2.2$ (right), where 
 $g=11$, ${\bar \phi}=0.25$, 
and $v_0 {\bar n}=6\times 10^{-4}$.  
 The solute part $\omega_{\rm sol}(\phi)$ 
is negligibly small for $\phi\ls 0.6$ here,  
but it grows abruptly for larger $\phi$ and 
  becomes equal to 
$\Delta\omega_0(\phi)$ at $\phi=\phi_\alpha$. 
This ensures the existence  of an interface 
connecting   phases $\alpha$ and $\beta$. 
}
\end{center}
\end{figure}

\begin{figure}[th]
\begin{center}
\includegraphics[scale=0.45]{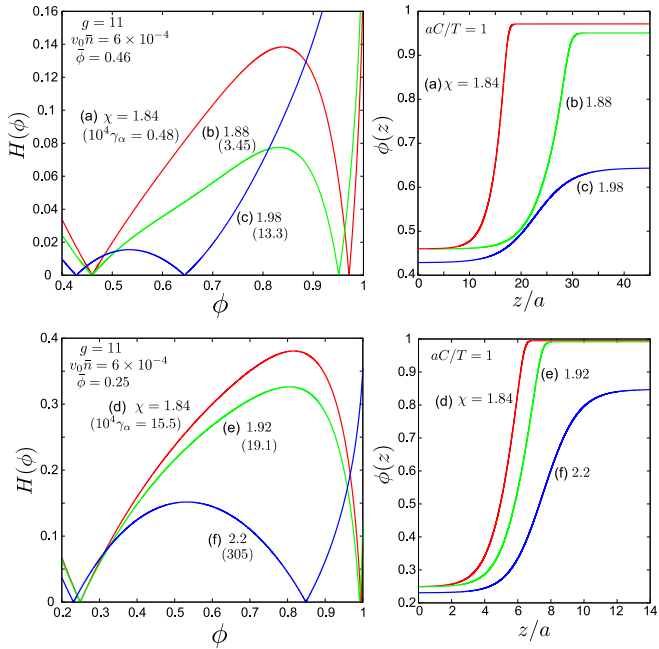}
\caption{Left:$H(\phi)$ in Eq.(2.56)
vs $\phi$,   which vanishes  at $\phi=\phi_\alpha$ 
and  $\phi_{\beta}$. 
Right: $\phi(z) $ vs $z/a$, 
where $a=v_0^{1/3}$ and $C= T/a$. 
Here  $g=11$, ${\bar \phi}=0.25$, 
and $v_0 {\bar n}=6\times 10^{-4}$.
In the upper plates 
three curves correspond to (a)
$\chi=1.98$, (b) $1.88$, and (c) 1.84, while 
in the lower plates 
we set  (d) $\chi=2.2$, (e) $1.92$, 
and (f) 1.84.  In the parentheses the volume 
fraction $\gamma_\alpha$ 
multiplied by $10^4$ 
is written for each curve (left). 
}
\end{center}
\end{figure}

In the Ginzburg-Landau scheme, 
the surface tension $\sigma$ is given by 
the $z$ integration of the (generalized) 
grand potential density $\Delta\omega(\phi)+ 
C(\phi')^2/2$  including   the gradient contribution. 
Use of Eq.(3.4) yields  
\be
\sigma  
=\int dz C (\phi')^2=
 \sqrt{CT/v_0} 
\int_{\phi_\beta}^{\phi_\alpha}d\phi H(\phi),
\en 
where we define the dimensionless function, 
\be 
H(\phi) = (2v_0/T)^{1/2} 
[\omega(\phi)-\omega(\phi_\beta)]^{1/2}.
\en
In our theory, $C$ is an arbitrary 
constant and $\sigma \propto C^{1/2}$.  
In Fig.8, we plot $H(\phi)$ and $\phi(z)$ 
for  (a) $\chi=1.84$, (b) 1.88, and (c) 1.98
  with $\bar{\phi}=0.46$ in the upper plates 
and for   (d) $\chi=1.84$, (e) $1.92$, and 
(f) 2.2 with  $\bar{\phi}=0.25$ in the lower plates.  
Here  $g=11$, $v_0{\bar n}= 
6\times 10^{-4}$, and  $C=T/a$. 
The values ($a^2\sigma/T$, ${\bar n}/n_{\rm p}$)  in these cases 
are (a) $(0.044,1.01)$, 
(b) $(0.024,1.17)$, 
(c) $(0.0021,2.25)$, 
(d) $(0.19,6.6)$, 
(e) $(0.16,7.7)$, and (f) $(0.064,\infty)$. 
In the case (c),  the system 
is close to the solvent criticality and 
$\xi_\beta$ is relatively long.  
In the case (f),  $\bar \phi$ is slightly 
larger  than the composition $0.249$ 
on  the oil-rich branch of the 
coexistence  curve without solute, 
so phase separation occurs even without solute 
or $n_{\rm p}=0$.  The shape of $H(\phi)$ in Fig.8  
is roughly triangular  except (f),  
so we obtain  a simple  estimate, 
\be 
\sigma  \sim {C}
(\Delta\phi)^2/{\xi_\beta} . 
\en 
Since $\xi_\alpha$ 
is considerably shorter than $\xi_\beta$,  
we may approximate the interface profile as 
 \be 
 \phi(z)-\phi_\beta \sim  
\Delta\phi \exp(-z/\xi_\beta),
\en 
in the region  $0<z \ls \xi_\beta$, neglecting 
the variation in the region $-\xi_\alpha\ls z<0$. 
 Substitution of Eq.(3.12)  into  
  Eq.(3.9) yields  Eq,(3.11). 
 The behaviors  (3.11) and (3.12) hold even 
in the  limit  $\gamma_\alpha \to 0$.

\subsection{Discontinuous  appearance of a 
droplet due to surface tension: minimum droplet radius $R_{\rm m}$}

As an example,  we  suppose   a single spherical droplet  
of phase $\alpha$ suspended in phase $\beta$ in equilibrium, 
accounting for  the effect of 
the surface tension $\sigma$. The droplet radius $R$ is so small 
that the following  condition holds: 
\be 
u\equiv \Psi(\phi_\alpha) 
\gamma_\alpha  \ll 1, 
\en  
where 
$ 
\gamma_\alpha = 4\pi R^3/3V.
$ and $\Psi(\phi_\alpha) \sim e^{g(1-{\bar\phi})}$. 
We add the surface free energy 
$ 4\pi \sigma R^2$ to 
$\Delta F$ in Eq.(2.36) and  expand the logarithmic factor 
there  as 
$\ln (1+u)\cong u-u^2/2$. The resultant 
total free energy reads 
\be 
\Delta F_{\rm tot} = V( -w\gamma_\alpha + \frac{1}{2}T{\bar n} 
u^2)  +4\pi \sigma R^2,
\en 
up to second order in $\gamma_\alpha$.
The coefficient $w$  is defined by  
\bea 
w&=& -\Omega (\phi_\alpha)+T{\bar n}\Psi(\phi_\alpha)\\
&=&  
\Omega (\phi_\alpha)( {\bar n}/n_{\rm p}-1)\\
&=&Tv_0^{-1}(\phi_\alpha-{\bar \phi})^2 
(\chi- \chi_{\rm p}),  
\ena 
where   $n_{\rm p}$ in the second line 
is  the minimum solute density in Eq.(2.46) 
and  $\chi_{\rm p}$ in the third line 
is  the minimum interaction parameter in Eq.(2.48). 
A droplet can exit  
only for $w>0$, which means 
 ${\bar n}>n_{\rm p}$ or  $\chi> \chi_{\rm p}$. 
It is convenient to introduce two characteristic lengths by 
\bea 
R_c&=&2\sigma/w,\\
R_{\rm m} &=&  (3\sigma V/2\pi T{\bar n})^{1/4}\Psi(\phi_\alpha)^{-1/2}, 
\ena
in terms of which  $ \Delta F_{\rm tot}$ in Eq.(3.14) 
is rewritten as 
\be 
 \Delta F_{\rm tot} / {4\pi \sigma}= R^2
- 2R^3/3R_c+ R^6/3R_{\rm m}^4 ,
\en 
For $\sigma\sim T/a^2$ we roughly 
estimate 
\bea 
R_c &\sim& a/( {\bar n}/n_{\rm p}-1)\sim a/
(\chi-\chi_{\rm p})
\nonumber\\ 
R_{\rm m} &\sim& (n_{\rm p}/{\bar n})^{1/4}(Va)^{1/4}  
e^{-g(1-{\bar\phi})/4}.  
\ena 
If ${\bar n}$ (or $\chi$) 
is increased   above  $n_{\rm p}$ (or $\chi_{\rm p}$),  
 $R_{\rm m}$ soon exceeds   $R_c$ for large $V$,   
 despite  the reducing 
  factor  $e^{-g(1-{\bar\phi})/4}$.

We should  minimize  $ \Delta F_{\rm tot}$ in Eq.(3.20) 
as a function of $R$. For $R>0$ 
 we require   $\Delta F_{\rm tot}<0$ 
and $\p \Delta F_{\rm tot}/\p R=0$.  The latter condition is 
written as 
\be 
1-R/R_c+ R^4/R_{\rm m}^4=0, 
\en  
under which  $\Delta F_{\rm tot}=4\pi R^2(2-R/R_c)/3<0$, 
so $R>2R_c$ is needed. This inequality 
can be satisfied only for  
\be 
R_c< R_{\rm m}/2 \quad {\rm or}\quad w>4\sigma/R_{\rm m},.
\en 
which will be evident  in Fig.18 below.  
If the above   condition  
is not satisfied, there is no equilibrium droplet. 
As $ R_{\rm m}/R_c \to 2$,  we have $R\to R_{\rm m}$. 
Here $u$ in Eq.(3.13) is of order 
$(n_{\rm p}/{\bar n})^{3/4}[a^3e^{g(1-{\bar \phi})}/V]^{1/4}$ 
and can be much smaller than unity for large $V$.
Thus $R_{\rm m}$ is the minimum equilibrium 
droplet radius in the presence 
of the surface tension. 
For $ R_{\rm m}/R_c\gg 1$, the surface free energy 
becomes negligible and 
\be 
R \cong (R_{\rm m}^4/R_c)^{1/3}= 
[3Vw/4\pi {\bar n}\Psi(\phi_\alpha)^2]^{1/3}. 
\en 
We also note that  the transition occurs for  
${\bar n}/n_{\rm p}-1\gs a/R_{\rm m}$ from Eq.(3.16) 
or for $(1-{\bar\phi})^2(\chi-\chi_{\rm p}) \gs a/R_{\rm m}$ 
 from Eq.(3.17).

\begin{figure}[t]
\begin{center}
\includegraphics[scale=0.44]{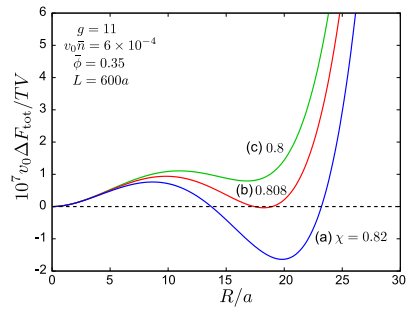}
\caption{(Color online) 
$ \Delta F_{\rm tot}$ including the surface free energy  vs 
$R/a$ for (a) $\chi=0.82$,  
(b) 0.808, and (c) 0.8 from below.  
At the minimum with positive $R$, the droplet 
is stable for (a), marginal 
 for (b), and metastable for (c),  
while  ${\bar n}/n_{\rm p}=1.080$ for (a), 
1.073 for (b), and 
1.067 for (c). Here  $R= 18.23a$ at the minimum for (b). 
}
\end{center}
\end{figure}
 
In Fig.9, we  plot  
$\Delta F_{\rm tot}=\Delta F+4\pi \sigma R^2$ as a function of $R$ 
in the spherically  symmetric geometry, 
where ${\bar\phi}=0.35$, 
$v_0{\bar n}=6\times 10^{-6}$,  and $g=11$. 
The cell volume is  $V=4\pi L^3/3
 \cong 0.9\times 10^9 v_0$ with $L=600a$.  
Here we use  Eq.(2.36) for $\Delta F$  
and not the expansion form of $\Delta F$  in Eq.(3.14). 
For the three curves in Fig.9, 
the parameter $u$ in Eq.(3.13) 
 is surely smaller than unity and 
Eq.(3.14) or Eq.(3.20) is a good approximation. 
In fact,  at the minimum position of  the curve (b) in Fig.9, 
we have  $R=18.23a$,  
$\sigma=0.231T/a^2$, 
 and $\Omega  =0.706T/v_0$, 
while $R_{\rm m} = 17.89a$ from Eq.(2.70) 
in this case.   In the next section, we will 
present simulation results in the 
same situation for hydrophilic ions. 

\subsection{Solvation-induced prewetting transition}

We note that Eq.(3.14) can also be used to examine 
the composition profile $\phi(z)$ near a boundary wall at $z=0$. 
Here, slightly below the precipitation curve 
$\chi=\chi_{\rm p}$ (or $\bar{n}=n_{\rm p}$), we may predict 
a first-order prewetting transition \cite{Cahn,PG} at 
$\chi=\chi_{\rm tr}(<\chi_p)$ with increasing $\chi$, 
where the surface adsorption changes discontinuously.   
This result is  related to the experiment by 
Beaglehole \cite{Be}.

Far from the wall, we assume that 
$\phi(z)$ and $n(z)$ tend to 
$\bar \phi$ and $\bar n$, respectively. 
Then Eq.(3.4) holds with 
\be 
\Delta \omega (\phi)= \Omega(\phi)- 
T{\bar n}\Phi(\phi),
\en 
where $\Omega(\phi)$ and $\Phi(\phi)$ are defined 
by Eqs.(2.37) and (2.38). The previous $\Delta\Omega(\phi)$ 
in Eqs.(3.4) and (3.5) becomes the above $\Delta \Omega(\phi)$ 
if  $\phi_\beta$ and $n_\beta$ there 
are replaced by ${\bar \phi}$ and $\bar n$, respectively. 
Below the precipitation curve   $\chi=\chi_{\rm p}$, 
 $\Delta\omega(\phi)$ is positive 
for $\phi>{\bar \phi}$ and 
again approaches zero at $\phi \cong \phi_\alpha$ 
 as $\chi \to \chi_{\rm p}$. See Fig.7 for its behavior. 
Assuming $\phi'(z)= d\phi(z)/dz<0$, 
we rewrite Eq.(3.4) as 
\be 
\frac{d\phi}{dz}= - \sqrt{2\Delta \omega (\phi)/C}.
\en 
In addition to the bulk free energy $F$ in Eq.(3.1), 
we assume the surface free energy, 
\be 
F_s= - \int dS \alpha_{\rm w} \phi =-S\alpha_{\rm w} \phi_s, 
\en 
where $\int dS$ is the surface integral at $z=0$, 
$S$ is the surface area, 
$\phi_s=\phi(0)$, 
and $\alpha_w $ is a parameter  
arising  from  the short-range 
interaction between the solvent and the 
 surface. Here $\alpha_w >0$ 
 if the boundary wall  is  hydrophilic 
 and the solute is hydrophilic.  
The boundary condition of $\phi(z)$ 
 at $z=$  is given by 
 \be 
 C \phi'(0)= - \alpha_{\rm w} .
\en 
Then  $\sqrt{2C \Delta \omega (\phi)}=\alpha_{\rm w}$ 
at $\phi=\phi_s$. The 
surface free energy density per unit area  
is then written as \cite{Cahn,PG} 
\bea 
\hspace{-3mm}
{\cal F}_{\rm ad} &=&
\int_0^\infty dz [\Omega(\phi)- T{\bar n}\Psi(\phi)] 
- \alpha_{\rm w} \phi_s \nonumber\\
\hspace{-3mm} &=&
\int_{\bar \phi}^{\phi_s} 
 d\phi  [\sqrt{2C\Delta\omega(\phi)} -\alpha_{\rm w}]
- \alpha_{\rm w}\bar{\phi}. 
\ena  
\begin{figure}[t]
\begin{center}
\includegraphics[scale=0.44]{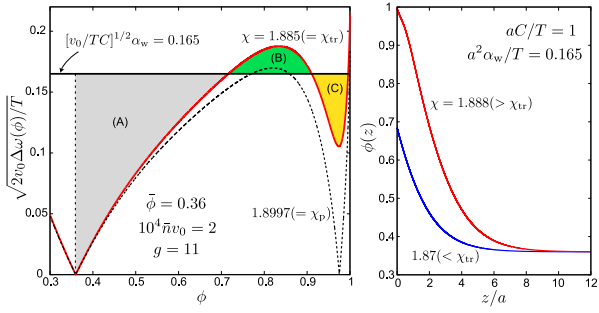}
\caption{(Color online) Left: $\sqrt{2v_0\Delta\omega(\phi)/T}$ 
vs $\phi$ with ${\bar\phi}=0.36$ and $v_0
{\bar n}=2\times 10^{-4}$, 
where $\Delta\omega(\phi)$ is defined in Eq.(3.25). 
For $\alpha_{\rm w}=0.165(CT/v_0)^{1/2}$, 
a first-order prewetting transition at a wall 
occurs at $\chi=\chi_{\rm tr}= 1.885$,  
where  the areas (B) and (C) coincide. 
Bulk two-phase coexistence is realized for 
$\chi>\chi_{\rm p}= 1.8997$. Right: $\phi(z)$ 
for $C=T/a$ 
near the wall before and after the prewetting transition.}
\end{center}
\end{figure}

Without  solute,  the composition increases 
 by    $\alpha_w \xi/C$ on the surface 
for $|\alpha_w| \ll T/a^2$, where $\xi$ 
is the correlation length. 
However, in the presence of solute 
with $g \gg 1$,  the adsorption on the 
surface can be strong with $\phi_s $ close to unity for 
$g\alpha_w  \xi/C \gs  1$ even when $\bar n$ is very small. 
In such cases,   $w_{\rm sol}$ 
and $\Delta w_{0}$ in Eq.(2.59) are  
of the same order on the surface (see Fig.7). 
In the left panel of Fig.10,  
 $\sqrt{2v_0\Delta\omega(\phi)/T}$ vs $\phi$ 
is displayed for $\chi=\chi_{\rm tr}= 1.885$ 
and $\chi_{\rm p}= 1.8997$, where  
 ${\bar \phi}=0.36$, 
${\bar n}=2\times 10^{-4}$, and 
 $\alpha_{\rm w}=0.165(CT/v_0)^{1/2}$. 
It demonstrates the presence of a first-order prewetting transition.  
In the right panel of Fig.10, 
the composition profile $\phi(z)$ 
is shown  slightly before 
and after the  transition.

\section{Hydrophilic salt }
\setcounter{equation}{0}

We now treat  aqueous  mixtures 
containing a small amount of   hydrophilic monovalent 
salt.  We shall see that  the  charge 
imbalance appears only near the interface as an electric double layer. 
In the statics of the present problem, the 
role of the electrostatic interaction    is  thus 
to shift  the surface tension  slightly 
(see the last paragraph of Sec.IVA)\cite{OnukiPRE}. 
In the dynamics, on the other hand, 
electric double layers 
around the droplet surfaces should suppress  fusion of  approaching 
droplets.  Such dynamical aspects are  
beyond the scope of this paper and should be studied in future.

\subsection{Preferential  ion solvation}

We write the  cation density as $n_1$ 
and the anion density as  $n_2$. Their 
 total amounts are fixed as 
\be 
\int d{\bi r} n_1= \int d{\bi r}n_2=V{\bar n}/2.
\en 
These densities both tend to 
$n_\alpha/2$ or $n_\beta/2$ 
in the bulk regions of phase $\alpha$ or $\beta$ 
due  to  the charge balance. 
The  electric charge density 
$e(n_1-n_2)$ gives rise to the electric 
potential $\Phi$ satisfying the Poisson equation, 
\be 
\nabla \cdot \ve \nabla\Phi= -4\pi e
(n_1-n_2).
\en 
The  dielectric constant 
$\ve(\phi)$ can depend on $\phi$ and 
has been assumed to be of the form,  
\be 
\ve(\phi)=\ve_0+ \ve_1\phi.
\en 
Neglecting  the image interaction   
\cite{OnukiPRE,Levin}, 
we assume 
the  total free energy $F$ in the following form,  
\cite{OnukiPRE}
\bea 
{F} &=& \int d{\bi r}\bigg[ 
{f(\phi)}  + \frac{1}{2}C|\nabla \phi|^2+ 
\frac{\ve}{8\pi }|{\nabla \Phi}|^2
\nonumber\\ 
&+&  T\sum_{i=1,2} [{n_i}\ln (n_i \lambda_i^3)- g_i\phi n_i ] \bigg ], 
\ena 
where the third term is the electrostatic 
contribution and the terms proportional to 
$g_i$ represent the solvation interaction among 
the ions and the solvent composition.
The $\lambda_1$ and $\lambda_2$ are the thermal de Broglie 
lengths of the two ion species.

Let us explain the solvation terms in more detail.  
The   ion   chemical potentials    
due to 	solvation in aqueous mixtures 
strongly depend on the ambient solvent composition 
$\phi$ \cite{Hung,OnukiPRE}. 
We write them as   $\mu_{\rm sol}^i(\phi)$, where $i$ 
represents the ion species. We   
assume  the linear form, 
 \be 
 \mu_{\rm sol}^i(\phi)= 
\mu_{0}^i- Tg_{i}\phi, 
\en   
where the first terms  are  irrelevant constants. 
This  linear dependence  is adopted to 
gain the physical consequences in the simplest manner and should 
not be taken too seriously. 
In aqueous solutions,   
 $g_{i}>0$   for hydrophilic  ions and 
 $g_{i}<0$ for hydrophobic ions.  
When two phases $\alpha$ and $\beta$ 
 coexist, the difference 
\be 
\Delta\mu_{\rm sol}^i =
 \mu_{\rm sol}^i(\phi_\beta)-\mu_{\rm sol}^i(\phi_\alpha)
\en   
 is called  the Gibbs transfer free energy 
in electrochemistry (usually measured 
in units of kJ$/$per mole), 
 leading  to an electric   potential difference across 
 an interface, called the Galvani potential difference.   
If we assume the linear form (3.5), it is expressed as 
\be 
 \Delta\mu_{\rm sol}^i= Tg_i\Delta \phi.
\en   
 For example, in  water-nitrobenzene(NB)  at  $300$K, 
 it can be estimated as 
 $17T$  for Na$^+$  and 
as  $19T$ for Cl$^-$ 
 per ion.  In this mixture, the two phases are strongly 
 segregated ($\Delta\phi \cong 1$) and  
 we  estimate  $g_i= 15$-$20$  for  
 small hydrophilic monovalent ions. 
In heavy water-tri-methylpyridine(3MP),  
the selective solvation 
has been found to induce mesoscopic phases 
with periodic structures 
for antagonistic salt. The Gibbs transfer free energies 
of  hydrophilic ions    for  water-3MP   
should not be larger 
 than  or  at least  of the same order as 
those for  water-NB, in view of the fact that   
 the  dielectric constants of 3MP and NB 
  are $9$  and  $35$, respectively.

We may 
calculate the composition and ion profiles  from  
the homogeneity of the chemical potentials 
$h= \delta F/\delta \phi$ 
and $\mu_i= \delta F/\delta n_i$, where 
\be 
h=f' -C\nabla^2\phi -\frac{\ve_1}{8\pi}|{\nabla \Phi}|^2 
- \sum_i g_i n_i.
\en 
If $\mu_i$ are homogeneous, we obtain  the modified 
Poisson-Boltzmann relations,
\be
n_i= \frac{1}{2}n_\beta 
 \exp [g_i(\phi-\phi_\beta) \mp  e(\Phi-\Phi_\beta)/T], 
\en  
where $-$ is for $i=1$ and $+$ is for $i=2$. 
The potential $\Phi$ tends  to constants 
$\Phi_\alpha$ and $\Phi_\beta$ in 
the  bulk regions. 
From the charge neutrality in the bulk regions, 
the potential difference between the two phases is given by 
\be 
\Delta\Phi= 
\Phi_\alpha-\Phi_\beta= T(g_1-g_2) \Delta\phi/2e.
\en  
Then  the ion densities 
in the two phases satisfy  
\be 
n_\alpha/n_\beta= \exp[(g_1+g_2)\Delta\phi/2].
\en 
The charge density appears only 
near the interface for $g_1\neq g_2$ forming an 
electric double layer (see Fig.11). 
As a result,  
the  bulk phase relations in Eqs.(2.8)-(2.18) 
for neutral particles  still hold for   
ions  with 
the correspondence relations, 
\be 
 n=n_1+n_2, \quad g=(g_1+g_2)/2.
\en  
For $g_1=g_2$ we have $n_1=n_2=n/2$. 
 
The thickness of the electric double layer  
is of the order of the Debye screening length,  
$\lambda_{\alpha}$ in the phase $\alpha$ side and 
$\lambda_{\beta}$ in the phase $\beta$ side, where  
\be 
\lambda_{K}=({
\varepsilon_K T/4\pi n_K e^2})^{1/2}, 
\en
where $K=\alpha$ or $\beta$.  For hydrophilic ions,  
$\lambda_{\alpha}\ll \lambda_{\beta}$ holds from   
$n_{\alpha}\gg n_{\beta}$ and 
$\lambda_{\alpha}$ is longer than the interface 
thickness except close to the solvent criticality. 
We assume  that 
the domain size is  much longer than 
  $\lambda_{\beta}$.

We found that the surface tension 
of an interface of a precipitated domain  
is slightly 
decreased by the electric double layer 
appearing for $g_1\neq g_2$. 
In fact, for $v_0{\bar n}\sim 10^{-3}$, 
it is larger  by a few $\%$ 
for  $g_1=15$ and $g_2=7$ 
than for  $g_1=g_2=11$, where 
 the other parameters 
are  common.

\subsection{Proposed experiment: a single droplet}

\begin{figure}[t]
\begin{center}
\includegraphics[scale=0.62]{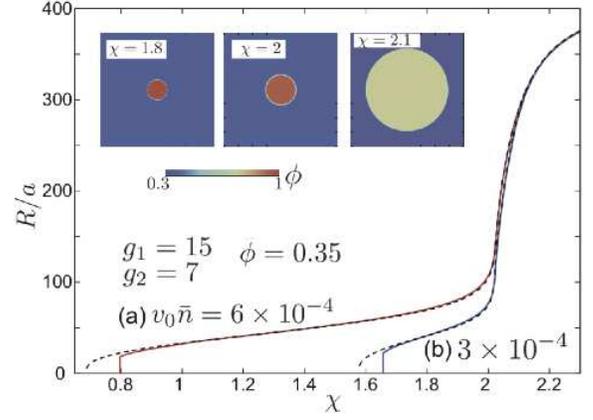}
\caption{(Color online)  
$R/a$ vs $\chi$ (bold line) 
with  hydrophilic ions 
calculated from  the free energy (4.4)  for (a) 
$v_0{\bar n}=3\times 10^{-4}$ (blue line) and 
(b) $6\times 10^{-4}$ (red line). 
Plotted also are theoretical curves from the approximate 
free energy (2.36) without the surface 
free energy (dotted line). 
Inset: Shapes of a spherical water-rich droplet 
in equilibrium 
for $\chi=1.8$, 2.0, and 2.1 
at $v_0{\bar n}=6\times 10^{-4}$,
in the region $-400<x,y<400$ and  $z=0$, where the colors 
represent the water composition according 
to the color bar. 
}
\end{center}
\end{figure}

\begin{figure}[t]
\begin{center}
\includegraphics[scale=0.48]{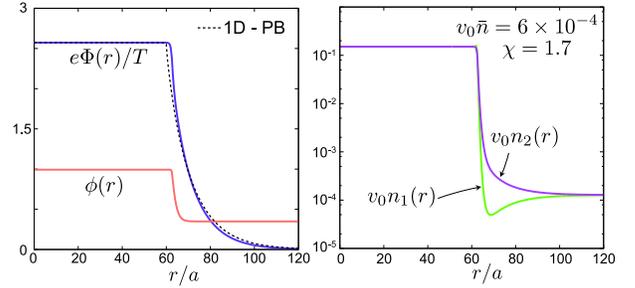}
\caption{(Color online) 
Normalized potential 
$e\Phi(r)/T$ (with $\Phi_\beta=0$) 
and water volume fraction 
$\phi(r)$ (left).  Normalized ion densities 
$v_0n_1(r)$ and $v_0n_2(r)$ (right). Here 
$\chi=1.7$  and $e\Delta\Phi=3T$. 
The other parameters are the same as in 
Fig.11.   In   the phase $\beta$ region near the interface, 
 $\Phi(r)$ well 
agrees with the one-dimensional 
solution of the nonlinear Poisson-Boltzmann 
 equation (dotted line) and 
 there appears an electric double layer 
 due to nonvanishing $n_1-n_2$.    
}
\end{center}
\end{figure}

We  performed simulation of 
 a   single water-rich droplet 
in the presence of hydrophilic ions.  
We  have already 
presented a theory for neutral solute in 
the same situation in Subsection IIF. 
Experimentally, wee may also   suppose 
a collection of monodisperse  droplets much separated 
from one another, where the volume per droplet (the inverse 
of the droplet density)  
should be treated as the system volume $V$.

The simulation details are as follows. 
We calculated equilibrium profiles 
of $\phi(r)$ and $n_i(r)$ by 
minimizing $F$ in Eq.(4.3) 
around a  spherical water-rich droplet with radius $R$ 
in  a spherical cell with radius $L=600a$.  
 At  $\bar\phi=0.35$, we set (a) ${\bar n}= 6\times 10^{-4}$ or  (b) 
$3\times 10^{-4}$.   The other parameters are 
  $g_1=15$, $g_2=7$,   
 $C= T/a$,  and  $e^2/ T=120a$. 
The dielectric constant depends on $\phi$ as 
 $\ve=40 (1+\phi)$. Since $g_1>g_2$,  an electric double 
 layer is produced at the interface  here.

In Fig.11, we show the equilibrium $R$ vs $\chi$ together with 
shapes of the droplet for three values of $\chi$.  
The droplet disappears due to the surface 
tension at $R=18.52a$ and $\chi= 0.799$ for (a)  and 
$R=22.18a$ and $\chi= 1.656$ for (b).  The corresponding 
value of ${\bar n}/{n_{\rm p}}$  is 1.067 for (a) and 1.102 for (b).
These curves are close to 
those  calculated from the approximate free energy 
(2.36) before the droplet disappearance. 
The characteristic features can also be calculated from 
the more approximate free energy (3.14) or (3.20).

In Fig.12, the profiles of 
the potential $\Phi(r)$ and the ion densities 
$n_1(r)$ and $n_2(r)$ are presented for $\chi=1.7$ in the case (a),  
which follow from Eqs.(3.2) and (3.7). 
Here  $\phi_\alpha 
=0.993$ and  $n_{\alpha}
= 0.352v_0^{-1}$ within the droplet  and  
 $\phi_\beta=0.349$ and  
$n_{\beta}=2.55\times 10^{-4}  v_0^{-1}$ 
outside it. The potential $\Phi(r)$ 
relaxes with the  Debye length  
$\lambda_\beta=  11.8 a$ and is well  
fitted to  the one-dimensional 
solution of the nonlinear Poisson-Boltzmann 
 equation (dotted line).   
In Fig.13, the equilibrium $R$ is displayed as a 
function of $\chi$ for (a) 
 $v_0{\bar n}= 6\times 10^{-4}$ 
and (b) $3 \times 10^{-4}$. Remarkably, 
the droplet disappeared at 
$R=18.52a$  and $\chi= 0.799$ for (a)  and 
at $R=22.18a$  and $\chi= 1.656$ for (b) due to the surface 
tension effect.   This critical radius 
for  (a) nearly coincides  with 
the droplet radius $R= 18.23a$ 
at the minimum of the curve (b) in Fig.9. 
We recognize that 
the relations of $R$ vs $\chi$ 
here  are little affected by 
the electric double layer. Hence  
they are close to 
those of neutral solvent 
with $g=11$ for  the same $\bar\phi$ 
and  $\bar n$.

\section{Homogeneous nucleation from  one-phase states}
\setcounter{equation}{0}

We   start  with 
homogeneous one-phase states without impurities 
(other than the solute or the salt under consideration). 
We assume that the boundary walls are hydrophobic 
and formation of a water-rich wetting layer is suppressed. 
In this situation,  
the  precipitation predicted in this paper   
can be  realized via homogeneous  nucleation for ${\bar n}>n_{\rm p}$ 
or $\chi>\chi_{\rm p}$. 
Therefore, we calculate the nucleation rate 
in the initial stage of nucleation. 
It is not much  affected by the electrostatic 
interaction for  hydrophilic ions  and 
we  treat neutral solute in the following. 
We also  neglect temperature inhomogeneity 
and treat $\chi$ as a homogeneous constant.

It is worth noting that 
water droplets  can easily be 
produced  around hydrophilic ions 
in metastable gas mixtures containing water vapor.
Here ions play the role of 
nucleation seeds on which hydration-induced 
 condensation is favored. 
A Ginzburg-Landau approach to this problem 
was also presented \cite{Kitamura}.

\subsection{Linear stability and metastability }

We first examine the linear stability 
of a one-phase state  
with $\av{\phi}= {\bar\phi}$ and $\av{n}
={\bar n}$. 
If $f_{\rm tot}$ in Eq.(2.3) 
is expanded with respect to the deviations 
$\delta\phi=\phi-{\bar \phi}$ and 
$\delta n=n-{\bar n}$, the second-order 
free energy  deviation  is written as  
\be
(\delta f_{\rm tot})_2
= \frac{1}{2} f''({\bar\phi}) (\delta\phi)^2 - 
Tg \delta\phi\delta n + \frac{T}{2{\bar n}} (\delta n)^2 .
\en  
If the right hand side  is 
nonnegative-definite, 
the one-phase  state is stable or metastable.  
 Minimization  with respect to $\delta n$ is achieved 
at $\delta n= g{\bar n}\delta\phi$, 
leading to $(\delta f_{\rm tot})_2 = 
[f''({\bar\phi}) - T{\bar n}g^2]
 (\delta\phi)^2/2$. 
Thus   the spinodal curve is expressed as 
\be 
f''(\bar{\phi})-T{\bar n}g^2=0.
\en  
See the left panels in Fig.1 for this  
spinodal curve. One-phase states are 
linearly stable  outside this curve.

In the presence of the Coulomb interaction, 
the spinodal  curve 
is still given by Eq.(5.2)  with 
the replacement $g=(g_1+g_2)/2$ in Eq.(4.12) 
under the condition $\gamma_{\rm p}<1$. 
Here $\gamma_{\rm p}$ is a  parameter 
representing  the degree of solvation asymmetry 
between the cations and anions (which should not be 
confused with the volume fraction $\gamma_\alpha$ 
of phase $\alpha$). For monovalent ions 
it is given by 
\be 
\gamma_{\rm p}= (T/16\pi C)^{1/2}|g_1-g_2|.
\en 
For  most hydrophilic ion pairs, 
we should have  $\gamma_{\rm p}<1$.
On the other hand, the reverse condition 
$\gamma_{\rm p}>1$ can be realized for 
antagonistic salt 
composed of hydrophilic and hydrophobic ion pairs 
\cite{Onuki-Kitamura,OnukiPRE,Sadakane}. 
Addition of such a salt 
leads to a decrease in 
the surface tension  and  mesophases formation. 
In this paper, we consider 
hydrophilic ions with $\gamma_{\rm p}<1$ 
for simplicity. 
 
\subsection{Droplet free energy and nucleation rate}

In the early stage of nucleation, the volume fraction 
$\gamma_\alpha$ of the new phase  is   very small and the 
droplets may be  treated  independently. 
Hence we consider a single droplet with radius $R$ in 
the classical nucleation theory \cite{Onukibook}. 
The composition and the solute density 
far from the droplet are  ${\bar\phi}$ and ${\bar n}$. 
A key quantity here is 
 the  free energy $\Delta F(R)$ needed 
to create a droplet  of the new phase.

To  derive  $\Delta F(R)$,  
we use  the total free energy density 
$f_{\rm tot}(\phi,n)$ in Eq.(2.3). 
It is equal to  $\bar{f}_{\rm tot} 
=f_{\rm tot}(\bar{\phi},\bar{n})$ 
in the initial homogeneous state and 
remains so far from the droplet, while it is equal to 
$f_{\rm tot}^\alpha 
=f_{\rm tot}(\phi_\alpha,n_\alpha)$ inside the droplet. 
Here  $\phi_\alpha$ and $n_\alpha$ 
are the composition and the solute density 
inside the droplet, respectively. 
The change of the total free energy is  
\bea 
&&\Delta F(R)= 
\frac{4\pi}{3}R^3 [f_{\rm tot}^\alpha-\bar{f}_{\rm tot}]
\nonumber\\
&&
+ \int_{r>R}
 d{\bi r} [f_{\rm tot}(\phi,n)-\bar{f}_{\rm tot}]  
 +4\pi\sigma R^2. 
\ena 
In the right hand side, the first term is the contribution 
from the droplet interior, the second term 
is that of the droplet exterior, and the last term 
is the surface tension term. 
Outside the droplet we assume small deviations 
of $\phi$ and $n$ from the initial values 
 $\bar \phi$ and $\bar n$ so that we set 
$f_{\rm tot}(\phi,n)-\bar{f}_{\rm tot}
= {\bar h}(\phi-{\bar \phi})+\bar{\mu}(n-\bar{n})$, 
where 
$\bar{h}$ and $\bar{\mu}$ are the values of $h$ in Eq.(2.6) and $\mu$ 
in Eq.(2.7) in the initial metastable state. 
Further we note that the integrals  
of $\phi-{\bar \phi}$ and $n-\bar{n}$ in the whole space 
vanish from the conservation relations (2.5), so that 
 their space integrals outside 
the droplet are equal to $-(4\pi/3)R^3 
 (\phi_\alpha-\bar{\phi})$ and 
$-(4\pi/3)R^3  (n_\alpha-\bar{n})$, respectively. 
Therefore, it follows 
the standard  form \cite{Onukibook}, 
\be 
\Delta F(R)= -\frac{4\pi}{3}R^3 w  +4\pi R^2 \sigma.
\en 
The coefficient $w$ represents the degree of metastability 
and is of the form, 
\bea
-w &=& f_{\rm tot}^\alpha-\bar{f}_{\rm tot}  
- \bar{h}(\phi_\alpha-{\bar\phi})
- \bar{\mu}(n_\alpha-{\bar n}) \nonumber\\
&=&\Omega(\phi_\alpha)-T{\bar n}\Psi(\phi_\alpha), 
\ena
where $\Omega(\phi_\alpha)$ and $\Psi(\phi_\alpha)$ 
are given in Eqs.(2.37) and (2.38). 
From the second line of Eq.(5.6), this $w$ coincides 
with $w$ introduced in Eq.(3.15).  
It is  proportional to 
${\bar n}-n_{\rm p}$ and $\chi-\chi_{\rm p}$ 
 as in Eqs.(3.16) and (3.17), where 
  $n_{\rm p}$ and $\chi_{\rm p}$ are defined 
  in  Eqs.(2.46) and (2.48).

The critical radius is $R_c=2\sigma/w$ as in Eq.(3.18).
At $R=R_c$, $\Delta F(R)$ attains a maximum 
given by 
\be 
F_c= 16\pi\sigma^3/3w^2. 
\en  
In the classical nucleation theory, 
critical  droplets appear with a constant 
rate $I$ per unit time and per unit volume. 
It is  called the nucleation rate and is  of the form, 
\be 
I=I_0 \exp(-F_c/T),
\en  
where the coefficient 
$I_0$ depends on the dynamics and will be estimated in Eq.(5.22). 
For weal metastability, 
the nucleation barrier   $F_c/T$ 
grows as $w^{-2}$ and $I$ is very sensitive to $w$. 
From Eqs.(5.7)  and  (5.8) 
the  nucleation barrier is expressed  as    
\bea
  F_c/T& =& A_c/(\bar{n}/n_{\rm p}-1)^2 \\
&=&  B_c /(\chi-\chi_{\rm p})^2.
\ena 
The two coefficients $A_c$ and $B_c$ are defined by    
\bea 
A_c &=&16\pi\sigma^3/3\Omega^2, \\
B_c&=&16\pi\sigma^3v_0^2/3T^2 (\phi_\alpha-{\bar \phi})^4 ,
\ena 
where   $A_c$ depends on $\chi$ and $\bar \phi$ 
and   $B_c$ on  $\bar n$ and $\bar \phi$.
We may use Eq.(5.9) or Eq.(5.10) 
depending on whether $\bar n$ or $\chi=\chi(T)$ 
is varied. 

\begin{figure}[t]
\begin{center}
\includegraphics[scale=0.44]{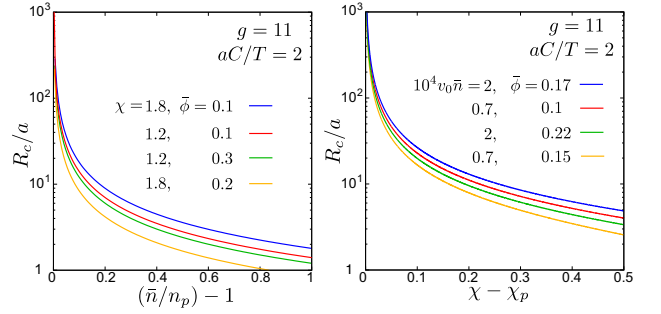}
\caption{(Color online) Critical radius $R_c$ in Eq.(3.18) 
 as a function of ${\bar n}/n_{\rm p}-1$ 
 for $(\chi,{\bar \phi})=(1.8,0.1), (1.2,0.1),(1.2,0.3), $ and 
 $(1.8,0.2)$ from  above (left) 
and  as a function 
of ${\chi}- \chi_{\rm p}$ for 
$(10^4v_0{\bar n},{\bar \phi})=
(2.0.17), (0.7,0.1),(2,0.22), $ and 
 $(0.7,0.15)$ from  above (right). 
Here  $g=11$ and $C=2T/a$. 
}
\end{center}
\end{figure}

\begin{figure}[t]
\begin{center}
\includegraphics[scale=0.44]{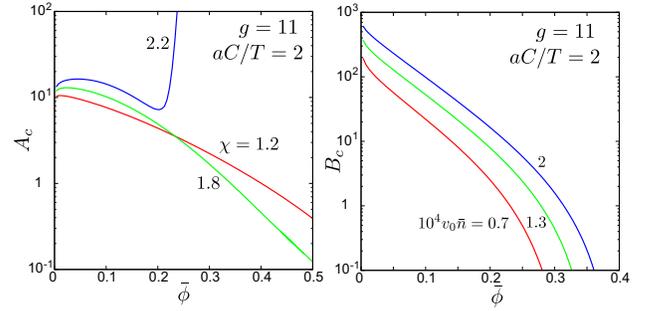}
\caption{(Color online) $A_c$ at fixed $\chi$ 
in Eq.(5.11) (left) and $B_c$ at fixed $\bar n$ 
in Eq.(5.12) (right) 
as functions of $\bar\phi$ 
for $g=11$ and $C=2T/a$. For $\chi>2$ (left),  
there is no barrier or $A_c\to \infty$ 
as $\bar\phi$ approaches the 
equilibrium coexistence composition. 
}
\end{center}
\end{figure}

In Fig.13, we plot $R_c$ vs 
 ${\bar n}/n_{\rm p}-1$ at fixed $\chi$ in the left 
and  $R_c$ vs 
${\chi}- \chi_{\rm p}$ at fixed $\bar n$ in the right. 
It  much exceeds the molecular size $a$ 
for weak metastability,  
where ${\bar n}/n_{\rm p}-1$ or ${\chi}- \chi_{\rm p}$ is small.   
In Fig.14, we plot $A_c$ at fixed $\chi$ 
and $B_c$ at fixed $\bar n$ as functions of $\bar \phi$ 
for $g=11$. These coefficients are considerably 
larger than unity for ${\bar \phi}\ls 0.2$, but 
become small with increasing $\bar \phi$. 
In these figures we set $C=2T/a$. 
Here notice the relations 
$R_c\propto C^{1/2}$, 
$\sigma \propto C^{1/2}$, and 
$F_c \propto C^{3/2}$. 
The coefficients $A_c$ and $B_c$ 
in Eqs.(5.11) and (5.12) 
depend on $C$ as $\sigma^3 \propto \xi^3 
\propto C^{3/2}$.  For aqueous fluids with  the 
hydrogen bonding network,  
a larger value of $C$ might be more 
appropriate  than in Figs.13 and 14.  
See the summary for more discussions on the choice of $C$.

Nucleation  experiments 
have been performed precisely 
on  near-critical binary mixtures 
(without  salt) \cite{Onukibook,Langer,Goldburg,Siebert}, 
where appreciable  droplets become  observable for 
$F_c/T \ls 50$.  From  Fig.14,  
we find  $F_c/T \ls  50$ for 
${\bar n}/n_{\rm p}\gs 1.2$  at ${\bar\phi}=0.1$ 
In our case, 
 once ${\bar n}>n_{\rm p}$  or $\chi>\chi_{\rm p}$, 
the nucleation rate should 
soon becomes  large enough 
for usual  observations of 
  droplets with  $R>R_c$.

\subsection{Weak metastability and droplet growth }

We may derive an approximate expression for $w$ 
in terms of $\bar \phi$ and $\bar n$ 
for  weak metastablilty. 
To this end, we consider an equilibrium  reference  state,  
where two bulk phases with $(\phi,n) 
=(\phi_\alpha,n_\alpha)$ and $(\phi_\beta,n_\beta)$ 
are separated by a planar interface. 
We assume that  the composition and the solute 
density ($\phi_\alpha, n_\alpha$)   
in the reference phase $\alpha$ are the same as those 
within the droplet. However, those $(\phi_\beta,n_\beta)$  
in the reference phase $\beta$ 
 are slightly  different from 
those $(\bar{\phi},\bar{n})$ 
in the initial metastable state. 
The  initial homogeneous  deviations are written as  
\be 
\delta{\bar \phi}={\bar \phi}-\phi_\beta, \quad 
\delta{\bar n}={\bar n}-n_\beta. 
\en 
From Eqs.(2.16) and (2.17) we obtain 
\be 
f_{\rm tot}^\alpha-f_{\rm tot}^\beta  
- {h_{\rm cx}}\Delta\phi
- {\mu_{\rm cx}}\Delta n=0,  
\en 
where $f_{\rm tot}^K 
=f_{\rm tot}(\phi_K,n_K)$ (with $K=\alpha, \beta)$). The 
 ${h_{\rm cx}}$ and ${n_{\rm cx}}$ 
 are the values of $h$ in Eq.(2.6) and $\mu$ 
in Eq.(2.7) in the reference  state (which are simply written as 
$h$ and $\mu$ in Sec.II). 
Further we note  the relation, 
\be  
\bar{f}_{\rm tot} - f_{\rm tot}^\beta 
- {h_{\rm cx}}\delta{\bar\phi}
- {\mu_{\rm cx}}\delta{\bar n} \cong 0,
\en 
which is valid to first order 
in $\delta{\bar\phi}$ and $\delta{\bar n}$. 
From  the first line of Eq.(5.6) we may eliminate 
 $f_{\rm tot}^K $ and $\bar{f}_{\rm tot}$ to obtain  
 the desired expression,
  \be 
w\cong 
 ( \bar{h}-h_{\rm cx}) \Delta\phi + (\bar{\mu}-\mu_{\rm cx}) \Delta n,  
\en 
where we have set $\phi_\alpha-{\bar\phi}\cong \Delta\phi$ 
and $n_\alpha-{\bar n}\cong \Delta n$.
From Eqs.(2.6) and (2.7) the initial deviations 
of the chemical potentials are 
expanded as 
\bea 
&&\bar{h}-h_{\rm cx}\cong f_\beta''\delta{\bar\phi} -Tg \delta{\bar n},
\nonumber\\
&&\bar{\mu}-\mu_{\rm cx}\cong 
T\delta{\bar n}/n_\beta -Tg \delta{\bar \phi},
\ena 
where $f''_\beta$ is equal to $\p^2f/\p \phi^2$ at $\phi=\phi_\beta$. 
Therefore, 
\be 
\frac{w}{T} 
\cong ( f_\beta''\frac{\Delta\phi}{T}-g \Delta n) 
\delta{\bar\phi} 
+( e^{g\Delta\phi}-1- g \Delta\phi)\delta{\bar n}.
\en

In Appendix B, we will derive 
the dynamic equation for the droplet radius $R(t)$ for weak metastability. 
It is of the standard form, 
\be 
\frac{\p R}{\p t}= \frac{\Lambda}{R}
\bigg( w-\frac{2\sigma}{R}\bigg), 
\en 
where $\Lambda$ 
is a kinetic coefficient defined in Eq.(B10) below. 
For dilute  solute, it  may be 
related to  the mutual diffusion constant 
of the mixture $D_m(\phi)$ by 
\be 
\Lambda = D_m(\bar\phi)/f''(\bar\phi)(\Delta\phi)^2.
\en 
For $R \cong R_c$ we have 
$\p R/\p t\cong \Gamma_c (R-R_c)$, where 
\be 
\Gamma_c= 2\sigma \Lambda /R_c^3.  
\en  
The inverse $\Gamma_c^{-1}$ 
is the time scale  of near-critical droplets. 
The coefficient $I_0$ in the nucleation rate 
(5.8) is determined by 
the droplet dynamics at $R \cong R_c$ (with appropriate 
thermal noises added)  and is  estimated as 
\cite{Onukibook}
\be 
I_0 \sim \Gamma_c /{\xi}^{3},
\en 
where $\xi$ is the correlation 
length in the initial metastable state.

For $g=0$ our results tend to those for 
incompressible binary mixtures  without salt  
\cite{Onukibook,Langer,Goldburg}. 
There, we use Eq.(5.22) and  
the relation $w-2\sigma/R  
=f_\beta''(\Delta{\phi})^2[\Delta_s-2d_0/R]$ 
in terms of  the supersaturation 
 $\Delta_s= \delta{\bar\phi}/\Delta \phi$ 
 and the capillary length 
 $d_0= \gamma/f_\beta''(\Delta{\phi})^2$. 

\section{Heterogeneous nucleation 
on hydrophilic colloid surfaces}
\setcounter{equation}{0}

In this section,  
we examine adsorption, wetting, and precipitation on 
a colloid surface. 
Here  the colloid surfaces are hydrophilic, while 
the  boundary walls are  hydrophobic 
as in the previous section. The colloid density $n_{\rm co}$ 
is so  small that  each colloid particle 
may be treated independently and the 
composition profile around it depends only on the 
distance $r$ from its  center. 
The effective cell volume 
for each particle  is 
$V=n_{\rm co}^{-1}$. 
For simplicity, we assume that the colloids are 
neutral without surface charge and 
the correlation length $\xi$ 
outside the colloids is shorter than the colloid radius $d$. 
 
In Sec.IIIC, we have already found  a  
 prewetting transition on a planar wall 
 for $\chi$ slightly below $\chi_{\rm p}$. 
Furthermore, we  shall see that  a wetting layer 
much thickens slightly 
above $\chi_{\rm p}$ due to precipitation.

\subsection{Simulation results: discontinuous and continuous 
transitions}

First, we present our  numerical results   
for a hydrophilic 
particle with radius  $d=15a$ or $25a$ 
placed at the center of a spherical cell 
with radius $L=10^3a$. 
The cell volume is $V=4\pi L^3/3$ and 
the colloid volume fraction is $\phi_{\rm col}= 
(d/L)^3$, so $\phi_{\rm col}=3.4\times 10^{-6}$ and $1.6\times 10^{-5}$ 
for $d=15a$ and  $25a$, respectively. 
We suppose hydrophilic ions using the model in Sec.IV, 
though  the electrostatic 
interaction  is not essential here. 
The other parameters used are  $\bar{\phi}=0.36$, 
$v_0n_0=2\times 10^{-4}$, 
$g_1=9$, $g_2=13$, $C=T/a$, 
$e^2/ T=120a$, and $\ve_1=\ve_0=40$. 
The correlation length $\xi$ far from 
the surface is shorter than $d$. 
In fact, it is $1.39a$ at $\chi=1.9$.

For the composition 
$\phi$ and the ion densities $n_1$ and $n_2$, 
the total free energy is $F_{\rm tot}= 
F+F_s$, where 
$F$ is given by  Eq.(4.3)  and 
$F_s$ is the surface free energy in Eq.(3.27).   
The composition $\phi(r)$ 
obeys Eq.(3.2) at a constant $h$. The 
 boundary conditions at $r=d$ and $L$ are given by  
\be 
C \phi'(d)=-\alpha_w, \quad \phi'(L)=0,
\en
where $\phi'(r)= d\phi(r)/dr$. 
The wetting parameter $\alpha_{\rm w}$ 
is set equal to $0.165 T/a^2$ as in Fig.10.

In Fig.15, we plot the preferential 
adsorption given by    
\be 
\Gamma= \int_{r>d} d{\bi r}[\phi(r)-\phi(L)], 
\en 
for  the two diameters $d=25a$ and $15a$. 
The prewetting transition discussed in Sec.IIIC 
occurs at $\chi=1.873$ for $d=25a$ 
and $1.893$ for $d=15a$. Remarkably, 
for $\chi>\chi_{\rm p}= 1.8997$, 
$\Gamma$ increases continuously for $d=25a$, 
but discontinuously for $d=15a$.
In Fig.16, we show the composition 
$\phi(r)$ for four  $\chi$ across the transitions 
and around the crossover. 
It changes from a thin to thick 
wetting layer  continuously (but abruptly) 
for $d=25a$ and  discontinuously for $d=15a$.   
In Fig.17, we show the free energy change   
$\Delta F_{\rm tot}= F_{\rm tot} - F_{\rm tot}^0$ 
as a function of $\chi$,  where $ F_{\rm tot}^0$ is 
the total free energy for the homogeneous state 
(realized for $\alpha_w =0$). 
Around the first-order transitions 
there appear two branches, where  Eq.(3.2) is satisfied, 
and  on   the equilibrium  branch 
$F_{\rm tot}$ takes a lower value. 
In particular, for $d=15a$,   two branches of 
  adsorption and precipitation appear around 
  $\chi=1.918>\chi_{\rm p}$.

The prewetting transition point 
  below $\chi_{\rm p} $  depends 
 on $d$ as in Fig.15. Furthermore, it is slightly 
 affected by  the 
 electric double layer for $g_1\neq n_2$. In fact, for 
 $g_1=g_2=11$, it disapperas and 
 the prewetting transition occurs 
at $\chi=1.907$ for $d=25a$ and   
at $\chi=1.912$ for $d=15a$,  while it was  
at $\chi=1.885$ for a planar wall in Fig.10.

\begin{figure}[t]
\begin{center}
\includegraphics[scale=0.5]{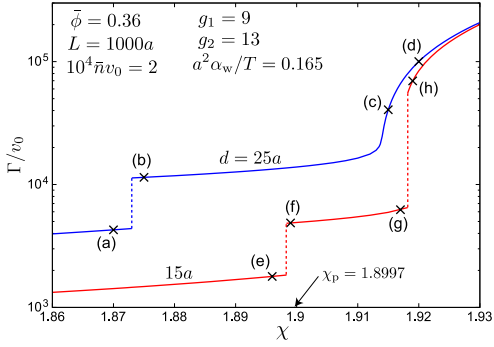}
\caption{Preferential adsorption $\Gamma$ 
on a colloid surface in Eq.(6.2) 
as a function of  $\chi$ on a semi-logarithmic scale 
at  $\bar{\phi}=0.36$.    
For $d=25a$ (upper curve), after a prewetting transition at 
$\chi=1.873$, the crossover to 
precipitation  is  continuous. 
For   $d=15a$ (lower curve),  after a prewetting transion at 
$\chi=1.898$, the transition 
to precipitation is discontinuous at $\chi=1.918$. 
At the points (a)-(h),  
the corresponding composition profiles are given in Fig.16. 
}
\end{center}
\end{figure}

\begin{figure}[t]
\begin{center}
\includegraphics[scale=0.4]{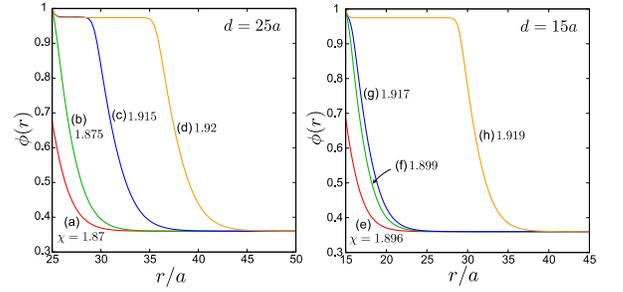}
\caption{(Color online)  Composition  $\phi(r)$ 
vs $r/a$ outside a  colloid $r>d$.  
For $d=25a$ (left), $\chi$ is (a) $1.87$,  (b) $1.875$, 
(c) $1.915$, and (d) $1.92$. For  
$d=15a$ (right), $\chi$ is (e) $1.896$,  (f) $1.899$, 
(g) $1.917$,  and (h) $1.919$. 
}
\end{center}
\end{figure}

\begin{figure}[t]
\begin{center}
\includegraphics[scale=0.44]{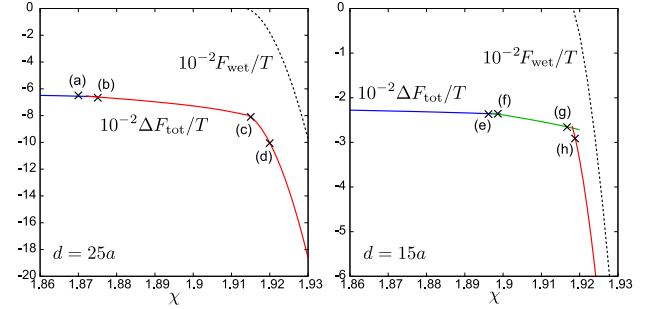}
\caption{(Color online) Free energy change 
$\Delta F_{\rm tot}$ due to 
water accumulation around a   colloid 
as  a  function of $\chi$  for ${\bar \phi}= 0.36$. 
For $d=25a$ (left), after the prewetting transion between (a) 
and (b), the crossover to precipitation  is  continuous 
but there is an abrupt slope change at (c). 
For   $d=15a$ (right)  after the  
 prewetting transion between (e) 
and (f), there is a discontinuous change between (g) 
and (h).  Shown also is the approximate 
free energy $ F_{\rm wet}$ in Eq.(6.6) 
(dotted line). 
}
\end{center}
\end{figure}

\subsection{Theory of wetting on a colloid surface}

 Before the layer thickening 
 due to precipitation and for $d\gg \xi$,  
$\Delta F_{\rm tot}$  is 
due to the adsorption 
expressed in terms of ${\cal F}_{\rm ad}$  
in Eq.(3.29) as  
\be 
\Delta F_{\rm tot}= 4\pi d^2 {\cal F}_{\rm ad}(\bar{\phi},\bar{n}), 
\en 
 We explicitly write the dependence of ${\cal F}_{\rm ad}$  
  on the composition $\bar \phi$ and the solute density $\bar n$ 
far from the surface. In Fig.17, ${\cal F}_{\rm ad}$ only weaky depends 
on $\chi$.  After precipitation in the range $\chi>\chi_{\rm p}$,  
a thick wetting layer appears in the region $d<r<R$ 
and $\Delta F_{\rm tot}$  is expressed as  
\be 
\Delta F_{\rm tot}= 
4\pi d^2 {\cal F}_{\rm ad}(\phi_\alpha,n_\alpha) 
+  F_{\rm wet}. 
\en  
The first adsorption term  is  
determined by the composition $\phi_\alpha$ and 
the solute density $ n_\alpha$ 
in the surrounding thick  layer, while the 
second one is  the contribution 
in the thick layer. Figure 17 
indicates  that ${F}_{\rm wet}$ 
decreases dramatically with increasing $\chi$.

To construct a  simple theory of   $F_{\rm wet}$,  
we assume the condition $\Psi(\phi_\alpha)\gamma_\alpha\ll 1$ 
in Eq.(3.13). We treat $\Delta F_{\rm tot}$ in Eq.(2.65) 
 as $F_{\rm wet}$. In the present situation,  
the volume fraction of phase $\alpha$ 
is given by 
\be 
\gamma_\alpha= 4\pi (R^3-d^3)/3V=(R^3-d^3)/L^3 .
\en 
Here $w$ in Eq.(3.15) is positive. 
In terms of the two lengths  $R_c\propto w^{-1}$  in Eq.(3.18) 
and $R_{\rm m}$  in Eq.(3.19),  we may express   $F_{\rm wet}$ as 
\bea 
&& \frac{ F_{\rm wet}}{ 4\pi \sigma}= R^2-d^2
- \frac{2}{3R_c}(R^3-d^3)+ \frac{1}{3R_{\rm m}^4}(R^3-d^3)^2 
\nonumber\\
&&\hspace{0.8cm}= 
{  R_{\rm m}^2}\bigg[  
 (q+ D^3)^{2/3}-D^{2} 
- \frac{2E}{3}q + \frac{1}{3}{q^2} \bigg].
\ena
The first line follows  from Eq.(3.20) if $R^2$ 
and $R^3$ there are replaced by $R^2-d^2$ 
and $R^3-d^3$, respectively.  
Thus, as  $d\to 0$, this $ F_{\rm wet}$ 
tends to $\Delta F_{\rm tot}$ in Eq.(3.20). 
In the second line we introduce the order parameter, 
\be
q =  (R^3-d^3)/R_{\rm m}^3,
\en 
where $D$ and $E$ are dimensionless parameters defined as 
\be 
D= d/R_{\rm m},\quad 
E= R_{\rm m}/R_c= wR_{\rm m}/2\sigma . 
\en 
We minimize $F_{\rm wet}$ as a function of $R$ or $q$. 
In Fig.17, this $F_{\rm wet}$ is plotted (in dotted lines) 
and is compared with $\Delta F_{\rm tot}$.  
In Fig.18, we display 
$q$ in the $D$-$E$ plane, which shows how 
a thick wetting layer appears. 
For  $D=0$ (or $d=0$) we have  the curve of 
$R^3/R_{\rm m}^3$ vs $R_{\rm m}/R_c$ for 
a droplet with radius $R$ 
(see Sec.IIF for its theory).

For $d-R \gg \xi$, 
 the adsorption free energy 
$4\pi d^2 {\cal F}_{\rm ad}$ is small 
as compared to $ F_{\rm wet}$. 
Then we may  examine  the formation 
of a thick  layer  by minimizing 
$ F_{\rm wet}$. 
For $q \ll D^3$, $ F_{\rm wet}$ is expanded in powers of $q$ as 
\be 
\frac{ F_{\rm wet}}{ 4\pi \sigma R_{\rm m}^2}=
\frac{2}{3}( \frac{1}{D}-E) q  
+ ({3} -  \frac{1}{D^{4}})\frac{q^2}{9} + \frac{4q^3}{81 D^{7}}  + \cdots
\en
By setting the coefficient of $q^2$ 
equal to zero, we find  a tricritical value of $D$ given by 
\be 
D_{\rm tri}=3^{-1/4}=0.760.
\en 
For $D=d/R_{\rm m} <D_{\rm tri}$, the second term ($\propto 
q^2$) in the expansion (6.9) is negative and 
 the transition takes place  
discontinuously 
on the transition line 
$E=E_{\rm tr}(D)$, where 
$E_{\rm tr} \to D_{\rm tri}^{-1}$ as $D \to D_{\rm tri}$ 
and $E_{\rm tr} \to 2$ as $D \to 0$.  
On the other hand, for $D=d/R_{\rm m} >D_{\rm tri}$, 
a thick wetting layer appears continuously as a 
second-order phase transition. 
The parameter  $q$ is  nonvanishing    for 
\be 
E> D^{-1}\quad {\rm or}\quad d>R_c, 
\en
where $q\propto (E-D^{-1})^{1/2}$. 
Thus  a tricritical point is 
at $(D,E)= (D_{\rm tri}, D_{\rm tri}^{-1})$ 
or at $(d,w)= (D_{\rm tri}R_{\rm m}, 
2\sigma /D_{\rm tri}R_{\rm m})$ 
for precipitation on a colloid surface.
In   Figs.15-17, 
we have $R_{\rm m}= 26a$, 
so that $D=0.96>D_{\rm tri}$ for 
$d=25a$ and $D=0.58<D_{\rm tri}$ for 
$d=15a$. Our numerical 
results are consistent with 
the  prediction from Eq.(6.6).

\begin{figure}[t]
\begin{center}
\includegraphics[scale=0.6]{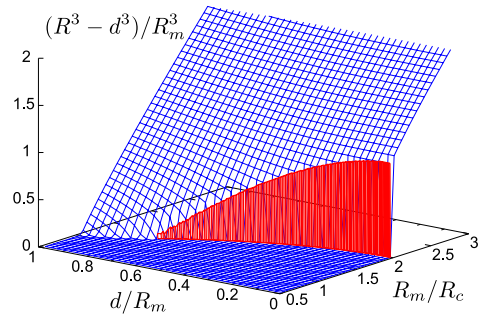}
\caption{(Color online) Precipitated 
layer volume $4\pi (R^3-d^3)/3$ 
divided by   $4\pi R_{\rm m}^3/3$ 
in the plane of $d/R_{\rm m}$ and $R_{\rm m}/R_c$. It is  
calculated from the approximate wetting 
free energy $F_{\rm wet}$ in Eq.(6.6) without 
the adsorption free energy $4\pi d^2{\cal F}_{\rm ad}$. 
The transition is discontinuous for $d/R_{\rm m}<0.760$ 
on the perpendicular surface (in red).  
}
\end{center}
\end{figure}

\section{Summary and Remarks}
\setcounter{equation}{0}

In summary, we have examined solvation-induced 
precipitation in aqueous mixtures with 
hydrophilic or hydrophobic solute 
in the limit of large $g$, which  
represents the 
composition-dependence of the 
solvation (the strength of the 
preferential solvation).

In Sec.II, we have developed a  thermodynamic 
theory of two-phase coexistence in mixture solvents with 
solute, numerical analysis of the two-phase coexistence, 
a  theory of precipitation 
in the asymptotic limit $g\gg 1$, and 
a theory in the dilute limit of 
precipitated domains. Remarkably, the precipitation 
curve $\chi=\chi_{\rm p}$ extends far below the 
coexistence curve in the $\chi$-$\bar \phi$ plane. 
In Sec.III, 
introducing the gradient free energy, we have 
calculated the surface tension $\sigma$,    
performed  stability analysis of a precipitated droplet 
including the surface tension, and 
predicted a prewetting transition below the precipitation 
curve. In Sec.IV, we have presented a theory 
for hydrophilic ions 
including the solvation and electrostatic interactions. 
Finally, we have investigated  homogeneous 
nucleation in Sec.V and 
heterogeneous nucleation on  a collod surface in Sec.VI. 
We  have also derived the evolution equation 
for the droplet radius $R(t)$ 
from the dynamic equations for $\phi$ and $n$ 
in metastable states in Appendix B.

We have presented a number of  predictions. They are 
(i) droplet appearance due to preferential 
solvation as in Figs.1-3, which has a minimim size 
$R_{\rm m}$ in Eq.(3.19), 
(ii) droplet size change  with varying $\chi$ or $T$ 
as in Fig.11,    (iii) prewetting transition 
slightly below the precipitation curve, 
(iv) nucleation slightly above the precipitation curve 
and far outside the solvent coexistence curve, 
 and (v) heterogeneous nucleation 
 on a collod surface, where 
 two-step first-order wetting transitions  
can occur as in Figs.15-18.
To confirm these effects, 
 systematic experiments are needed, 
where  the temperature, the water volume fraction, 
and the  salt amount are controlled. 
Particularly, dynamic light scattering is informative 
to detect emergence of droplets 
and wetting on colloid surfaces. 
Wetted colloids should move with a smaller diffusion 
constant $\propto R^{-1}$.    

In our theory,  the molecular volumes 
of the two components are assumed to be 
given by the common $v_0$ in Eq.(2.4). 
However, they can be very different. 
For example, those  of 
 D$_{2}$O and 3MP (the inverse densities of 
the pure components) are 28 and 
 168 \AA$^3$, respectively. 
Moreover, the coefficient $C$ in Eq.(3.1) 
of the gradient free energy 
remains  an arbitrary constant and the nucleation rate 
calculated in Sec.V is proportional to 
$C^{3/2}$. Therefore, our theory is still very  
qualitative. 
For pure water, 
the observed surface tension  outside the critical 
region  $1-T/T_c \gs 0.1$ 
 fairly agrees with the calculated  surface tension 
from the van der Waals  model  with 
the gradient free energy density 
$10 Ta^5 |\nabla n|^2/2$ \cite{Kitamura,Kiselev}, 
where $n$ is the density 
and $a=3.1{\rm \AA}$ 
is the van der Waals radius.

In future, we  will  investigate  the 
wetting transition on  charged walls, rods, 
and  colloids and the solvation-induced colloid 
interaction, which can be 
 much influenced  by the ion-induced 
precipitation. The degree of ionization 
should also be treated as a fluctuating variable 
sensitively depending on the local environment \cite{Okamoto}.

\begin{acknowledgments}
One of the authors (A.O.) 
would like to thank M.A. Anisimov 
for valuable discussions in an early stage 
of this work. This work was supported by Grant-in-Aid 
for Scientific Research on Priority 
Area ``Soft Matter Physics'' from the Ministry of Education, 
Culture, Sports, Science and Technology of Japan.
\end{acknowledgments}

\vspace{2mm} 
{\bf Appendix A: Two-phase coexistence for $\chi>2$ 
at very small $\bar n$}\\
\setcounter{equation}{0}
\renewcommand{\theequation}{A\arabic{equation}}

For $\chi>2$ two-phase coexistence is possible without 
solute.  Let $\phi_\alpha^0$ and  $\phi_\beta^0$ 
be the equilibrium volume fractions 
in the two phases without solute. 
Here we calculate the deviations of the volume fractions 
$\delta \phi_\alpha=  \phi_\alpha
-  \phi_\alpha^0$ and $\delta \phi_\beta=  \phi_\beta 
-  \phi_\beta^0$ to linear order in the  solute density $\bar n$ 
in the mean field theory. We show that the linear theory 
holds  only  for very small $\bar n$.

The $n_\alpha$ and $n_\beta$ are still given by Eqs.(2.8) 
and (2.9).  From Eqs.(2.15) and (2.17) 
 the chemical potential difference  $h$ is expanded as 
$h=-T \Delta n/\Delta \phi$ so that  
\bea 
&&\delta \phi_\alpha = Tn_\alpha (g\Delta \phi -1+  e^{-g\Delta \phi})
/ f''\Delta\phi,\nonumber\\
&&\delta \phi_\beta = Tn_\beta (g\Delta \phi +1-  e^{g\Delta \phi})
/ f''\Delta\phi,
\ena 
where   $f''= f''(\phi_\alpha^0)= f''(\phi_\beta^0)$ 
for the symmetric free energy density in Eq.(2.4). 
Here $f''(\phi)= \p^2f/\p \phi^2$.

As the solvent critical point ($\chi_c=2$) 
is approached, the inequality 
$\Delta \phi\ll 1/g$ holds eventually to give  
\be 
\delta \phi_\alpha \cong - 
\delta \phi_\beta \cong 
T{\bar n}g^2\Delta\phi/2f''. 
\en 
If we set $\epsilon=\chi-2(\propto 1-T/T_c$), 
the above deviation is of order $v_0{\bar n} \epsilon^{-1/2}$. 
The  linear approximation holds when $\delta\phi_\alpha 
 \ll \phi_\alpha^0-1/2 \cong \Delta\phi/2$. Thus 
 the upper bound of the linear regime 
 of the solute doping  is very small as  
\be 
{\bar n}\ll  \epsilon^{3/2}/g^2v_0.
\en 
In the left panels of Fig.1, $\phi_\alpha-\phi_\beta $ 
is not small even for $\chi \cong 2$ 
due to the nonlinear solute effect.

Away from the critical point, 
the relation $g\Delta \phi\gg 1$ holds 
for large $g$. In this case, we find  
\be 
\delta \phi_\alpha \cong Tgn_\alpha/f'',\quad 
\delta \phi_\beta \cong -Tn_\alpha/f'' \Delta\phi, 
\en  
so that 
 $\delta \phi_\alpha
  \cong -(g\Delta\phi) \delta \phi_\beta>0$. 
From Eq.(2.13)  $\delta\phi_\alpha (\propto n_\alpha)$ 
increases steeply as ${\bar \phi}$ 
is decreased to $\phi_\beta^0$ (or as $\gamma_\alpha$ 
is decreased). This tendency can be seen in the left panels of Fig.1. 
The maximum of $n_\alpha$ is ${\bar n}e^{g\Delta\phi}$. 
The condition that  $\delta \phi_\alpha$ is much 
smaller than unity is written as 
\be 
{\bar n}\ll f''e^{-g\Delta\phi}/Tg
\en   
If  Eq.(A3) or Eq.(A5) does not hold,   $\phi_\alpha$  
nonlinearly deviates from $\phi_\alpha^0$ 
with respect to $\bar n$.

\vspace{2mm} 
{\bf Appendix B: Droplet growth }\\
\setcounter{equation}{0}
\renewcommand{\theequation}{B\arabic{equation}}

For weak metastability, 
 we derive the droplet-evolution equation 
(5.19)  in the neutral solute case.
The simplest dynamic equations 
for the composition $\phi({\bi r},t)$ 
and the solute density $n({\bi r},t)$ are given by   
the diffusive equations, 
\bea 
&&
\frac{\p \phi}{\p t}= L \nabla^2 (h-C \nabla^2\phi),\\ 
&&\frac{\p n}{\p t}= D_s  \nabla \cdot n\nabla\cdot (\mu/T), 
\ena
where $L$ is the kinetic coefficient for the composition, 
$D_s$ is the solute diffusion constant, 
and  $h$ and $\mu$ are given in Eqs.(2.6) and (2.7). 
Around a spherical droplet with radius $R$, 
all the quantities depend on the distance $r$ from 
the droplet center and the time $t$.

Slightly outside the droplet surface $r-R \gs \xi$,  
the gradient term in $h$ is negligible, where $\xi$ is 
the interface thickness. When the droplet 
growth or shrinkage is slow, 
we may use the quasi-static approximation:   
\bea 
&&h\cong \bar{h} + (h_R -\bar{h})R/r,\nonumber\\ 
&&\mu \cong \bar{\mu} + (\mu_R -\bar{\mu})R/r,
\ena 
where $h$ and $\mu$ tend to $\bar h$ and $\bar \mu$ 
far from the droplet and to $h_R$ and $\mu_R$ 
near the droplet surface. 
The conservations of $\phi$ and $n$ at the interface yield 
\bea 
&&\frac{\p R}{\p t}\Delta\phi= \frac{L}{R}(h_R -\bar{h}), \nonumber\\
&&\frac{\p R}{\p t}\Delta n= {\bar n}\frac{D_s}{R}(\mu_R -\bar\mu),
\ena 
where $\Delta\phi=\phi_\alpha-\phi_\beta$ and 
$\Delta n=n_\alpha-n_\beta$ as in Eq.(2.13).

In the interface region $|r-R|\ls \xi$, 
the generalized chemical potential 
$\mu -C\nabla^2\phi$ including the gradient term 
may be treated as a constant $h_R$,  so that 
\be 
h_R = h -C(\frac{\p^2}{\p r^2} + \frac{2}{r}\frac{\p}{\p r})\phi .
\en 
The solute chemical potential $\mu$ is also 
a constant $\mu_R$ near the interface. 
We set $\phi= \phi_{\rm int}(r-R)+\delta\phi $ and 
$n= n_{\rm int}(r-R)+\delta n $ in the interface region. 
Here $ \phi_{\rm int}(z)$ is 
the  one-dimensional 
interface solution of Eq.(3.4) 
and $n_{\rm int}(z)=n_\beta e^{g(\phi_{\rm int}(z)-\phi_\beta)}$. 
These represent the reference profiles 
and the  corresponding $h$ and $\mu$ are written as 
$h_{\rm cx}$ and $\mu_{\rm cx}$ as in Eq.(5.14). 
To linear order in $\delta\phi$ and $\delta n$, 
we obtain $\mu_R-\mu_{\rm cx}=
T \delta n/n_{\rm int}-Tg \delta\phi$ or 
\be 
\delta n= n_{\rm int} 
[(\mu_R-\mu_{\rm cx})/T + g \delta\phi], 
\en 
in the  region $|r-R|\ls \xi$.  We may linearize Eq.(B5) as 
\bea 
&&h_R-h_{\rm cx}= \bigg[f''(\phi_{\rm int})
-Tg^2 n_{\rm int}- 
 C\frac{\p^2}{\p r^2}\bigg]\delta\phi\nonumber\\ 
&& -\frac{2}{R}C\phi'_{\rm int} - g(\mu_R-\mu_{\rm cx})n_{\rm int}, 
\ena 
Then we  multiply Eq.(B7)  
by $\phi_{\rm int}'= d\phi_{\rm int}(r-R)/dr$ 
 and integrate over the 
 region $|r-R|\ls \xi$ to obtain 
\be 
(h_R-h_{\rm cx})\Delta\phi\cong 
\frac{2\sigma}{R}  -(\mu_R-\mu_{\rm cx})\Delta n,
\en 
where the first term in the right hand side of 
Eq.(B7) does not contribute 
and use has been made of Eq.(3.9) and 
$gn_{\rm int}\phi_{\rm int}'= dn_{\rm int}/dr$.
We rewrite Eq.(B8) as  
\be 
(h_R-\bar{h})\Delta\phi+  
(\mu_R-\bar{\mu})\Delta n
\cong\frac{2\sigma}{R}-w,  
\en 
where $w$ is given in Eq.(5.16) in the linear form. 
Using   Eq.(B4) we eliminate $h_R-\bar{h}$ 
and $\mu_R-\bar{\mu}$ from Eq.(B9)  
to obtain  the droplet evolution equation (5.19) with 
\be 
{1}/{\Lambda}= (\Delta\phi)^2/L +(\Delta n)^2/D_s{\bar n}.
\en 
where the second term in the right hand side is negligible 
for  dilute solute, leading to $\Lambda\cong L/(\Delta\phi)^2$ 
and Eq.(5.20).


\begin{thebibliography}{99}
\bibitem{SCAT} The dynamic  scattering intensity  
$G(q,t)$ consists  of the contribution from the composition fluctuations 
and that from the  heterogeneities. 
The latter is proportional to 
$\phi_{\rm h}R^3\exp(-D_Rq^2t)$ for  
 wave number $q$. Here the heterogeneities have a  
radius $R \ll 2\pi/q$ 
at a volume fraction $\phi_{\rm h}\ll 1$.
The diffusion constant $D_R$ is given by 
the Stokes formula ($\propto R^{-1})$.
\end{thebibliography}
\end{document}